\documentclass[iop]{emulateapj}
\usepackage{color}

\newcommand{\mathN}{\hbox{$\mathcal{N}$}}

\newcommand{\irac}{$3.6\mu m-4.5\mu m$}
\newcommand{\chone}{$3.6\mu m$}
\newcommand{\chtwo}{$4.5\mu m$}

\newcommand{\add}[1]{{\color{black} #1}}

\shorttitle{Ultradeep IRAC imaging over the HUDF and GOODS-S}
\shortauthors{I. Labb\'e et al.}

\begin{document}
\def\figlayout{
\begin{figure*}
\begin{center}
$$\includegraphics[height=10.5cm,trim=5.cm 0 0cm 2cm]{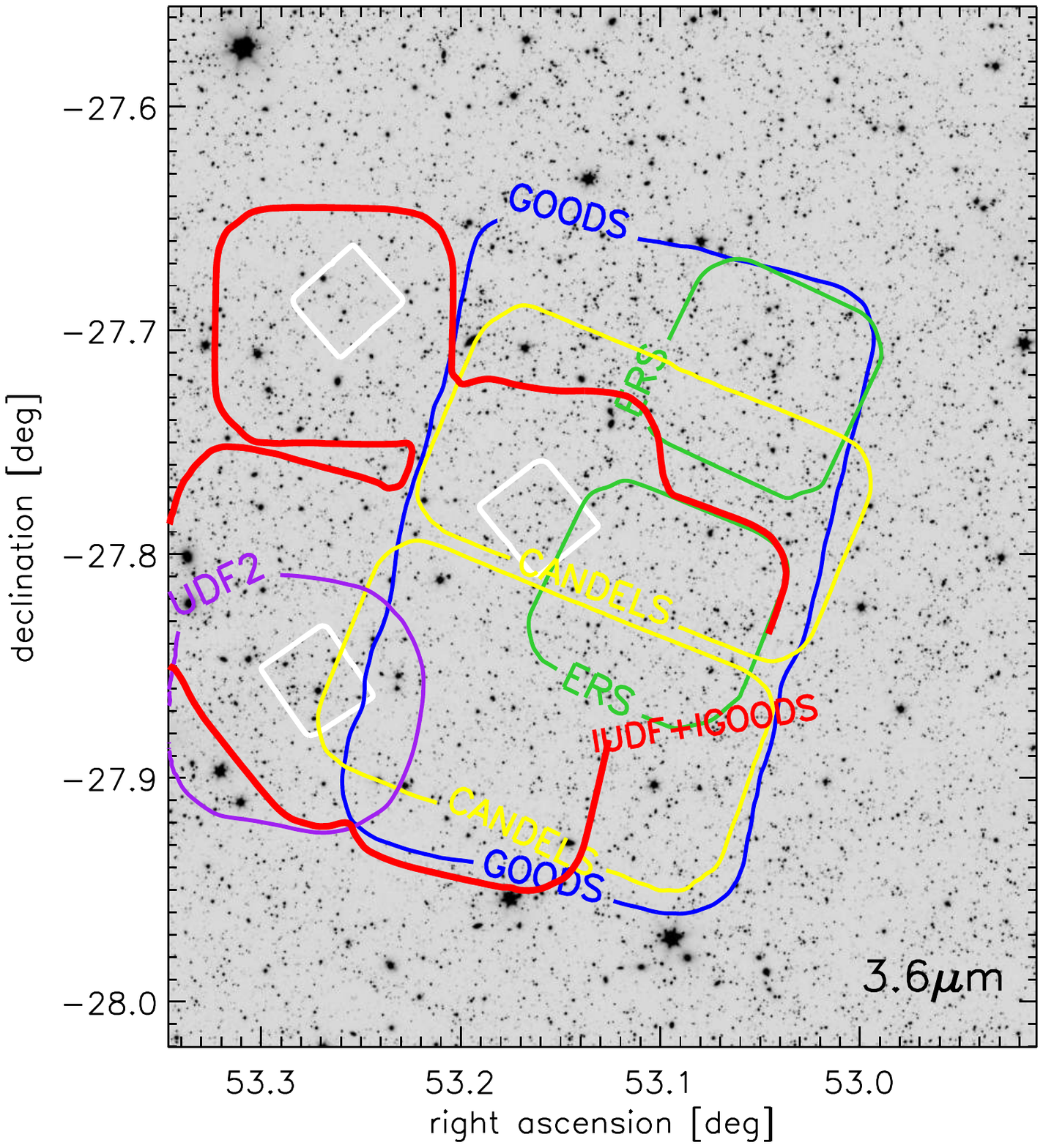}
\includegraphics[height=10.5cm,trim=10.55cm 0 0  2cm]{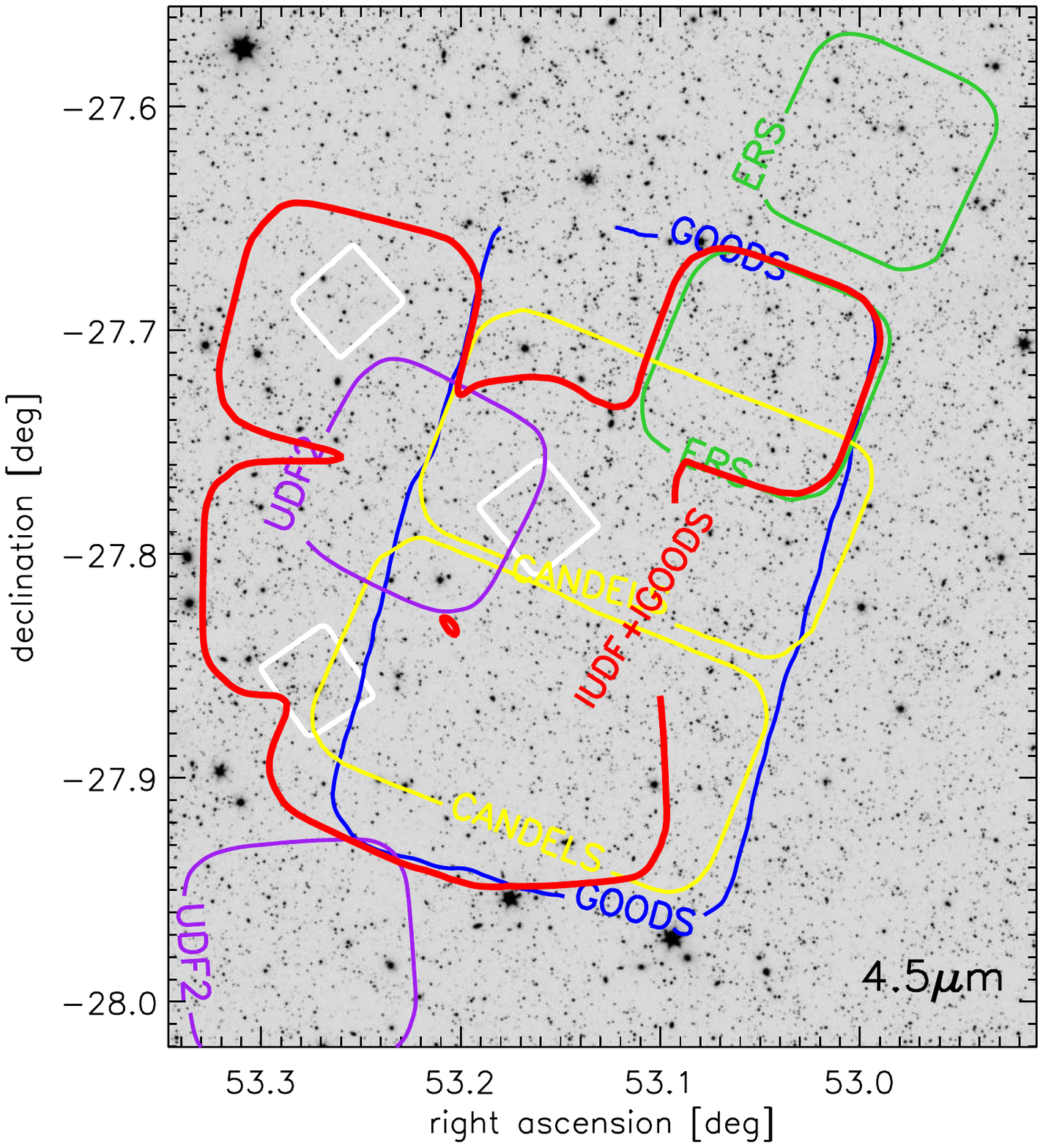}$$
\end{center}
\vspace{-0.7cm}
\caption{
Layout of the IUDF and IGOODS observations ({\em red}) on top of the IRAC imaging  at \chone\ ({\em left}) and \chtwo\ ({\em right}) from SIMPLE (Damen et al. 2011).  Also shown are all other ultradeep IRAC observations used in this paper, including warm mission data from ERS ({\em green}), S-CANDELS ({\em yellow}), and cryogenic data from GOODS ({\em blue}) and UDF2 ({\em purple}). Table 1 lists the all programs and PIs. The IUDF observations cover the HUDF/XDF and the two parallel fields ({\em white}), while IGOODS fills out part of the GOODS-South area.
\label{fig:layout}}
\vspace{0.4cm}
\end{figure*}
}

\def\figpsf{
\begin{figure}
\begin{center}
\includegraphics[height=8.3cm,trim=0.4cm 0 0cm 0]{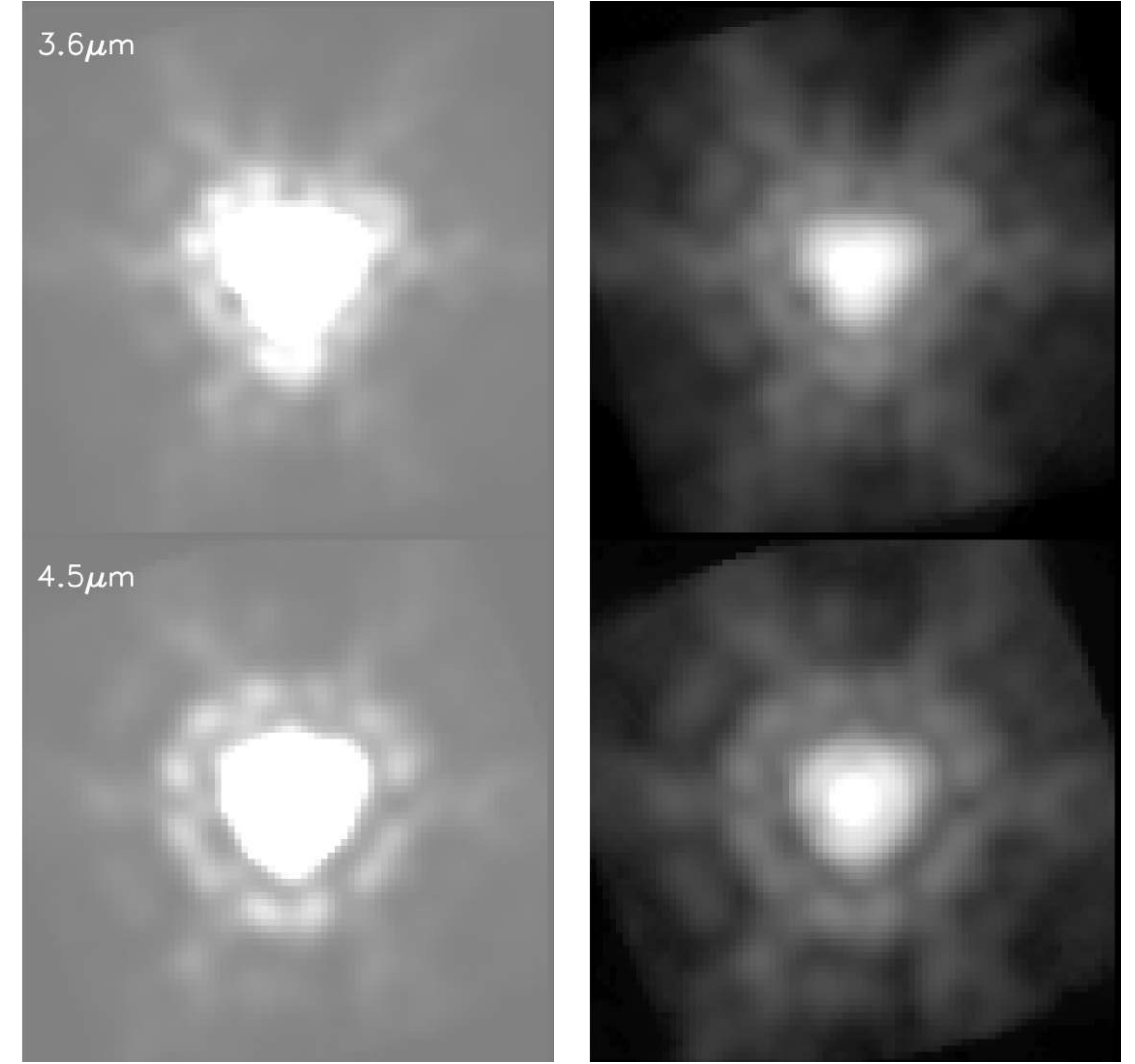}
\end{center}
\vspace{0.0cm}
\caption{
The empirical template PSF at \chone\ and \chtwo\ created from stacked images of stars spread
across the field from all 353 AORs. The left column shows the PSFs with linear scaling, the right column with a logarithmic scaling to capture the entire dynamic range and highlight the core structure as well as the PSF wings. The images are $24\farcs4\times24\farcs4$ and the PSF is sampled on a $0.3\arcsec$ grid ($\sim1/4$th native IRAC pixel).
\label{fig:psf}}
\vspace{0.5cm}
\end{figure}
}

\def\figpsfmap{
\begin{figure}
\begin{center}
\includegraphics[width=8.5cm,trim=0cm 0 0cm 0]{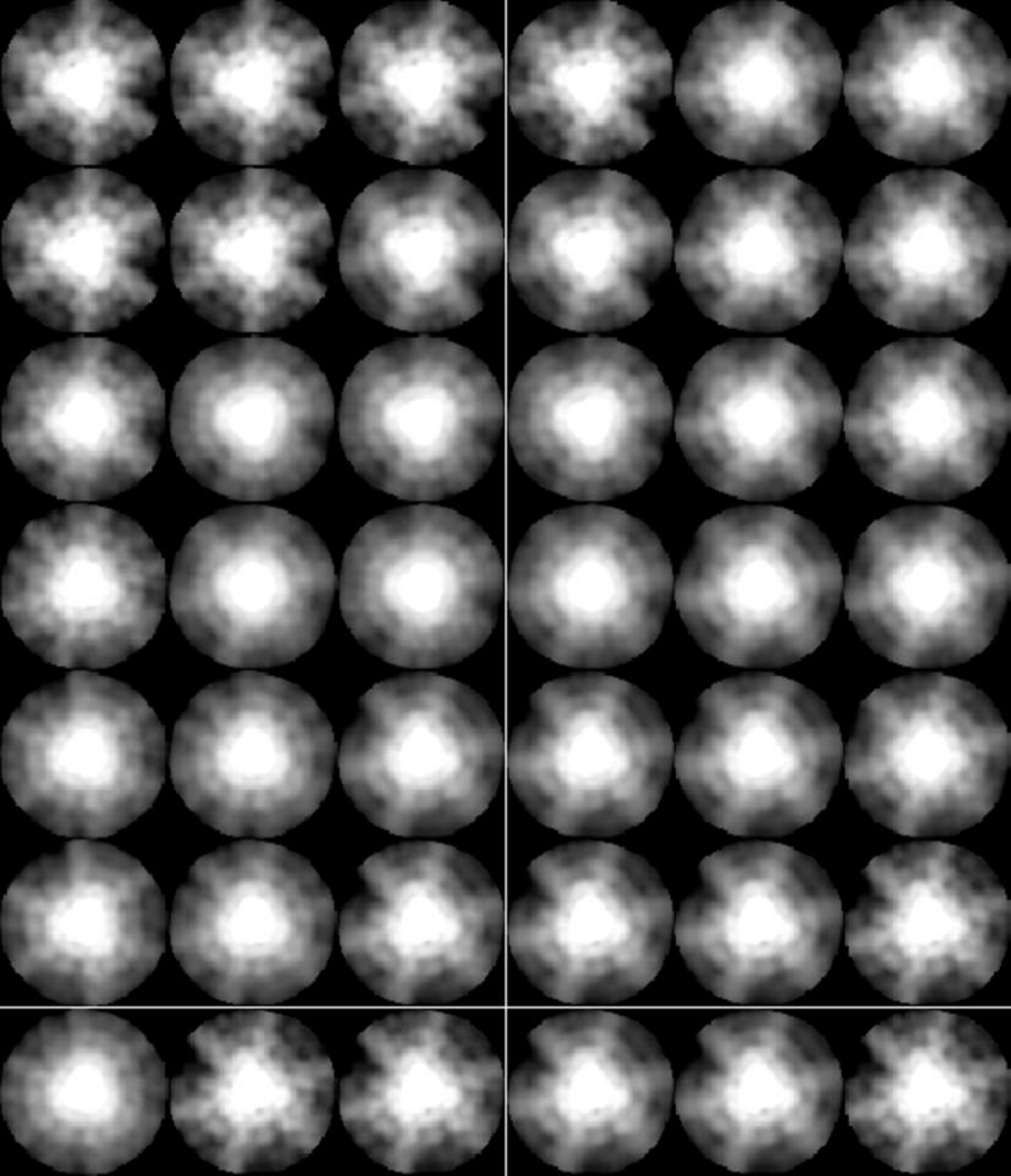}
\end{center}
\vspace{0.0cm}
\caption{
 The reconstructed 3.6\micron\ PSF mapped on a coarse grid in steps of $2.5\arcmin$,
 highlighting the spatial variation over the $12.5\arcmin \times 15\arcmin$ central
 area. The PSFs map is created by rotating and combining the template PSFs in the
 same way as the science data.
\label{fig:psfmap}}
\vspace{0.4cm}
\end{figure}
}

\def\figmosaic{
\begin{figure*}
\begin{center}
\vspace{-1.5cm}
$$\includegraphics[height=11.5cm,trim=5.2cm 0 0cm 0]{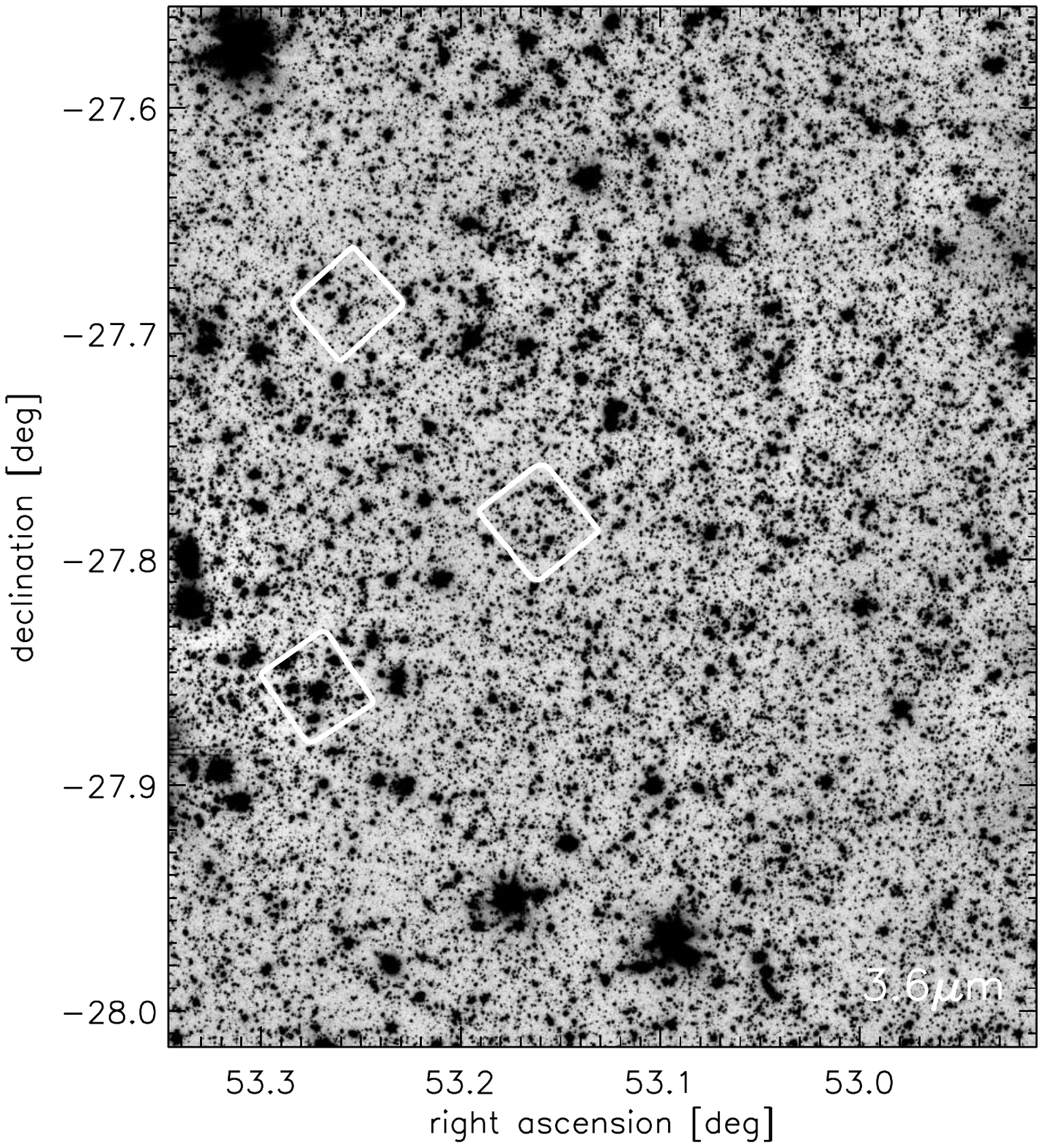}
\includegraphics[height=11.5cm,trim=10.3cm 0 0 0]{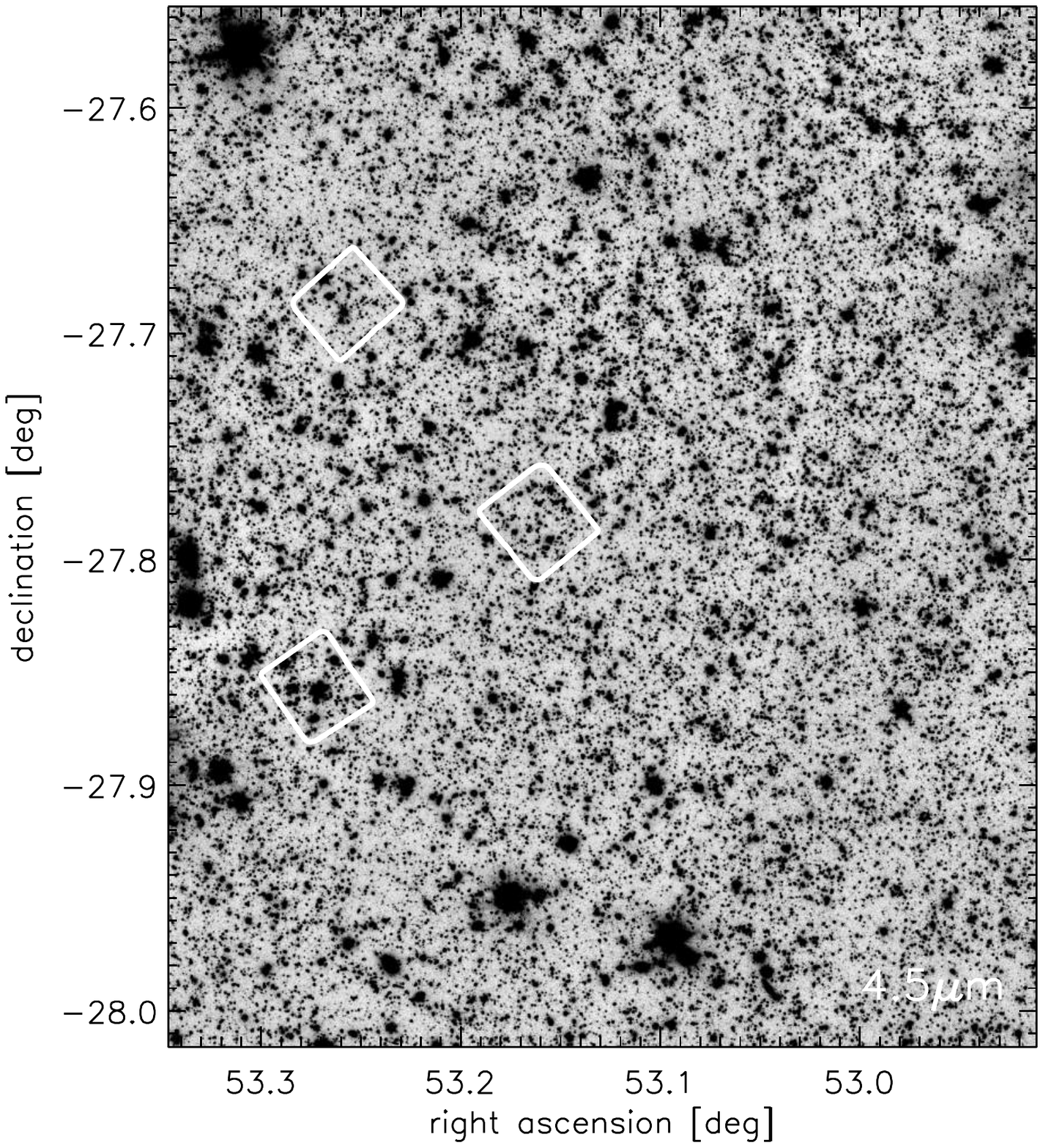}$$
\end{center}
\vspace{-0.7cm}
\caption{
 The full IRAC mosaics over GOODS-South and the HUDFs at \chone\ ({\em left}) and \chtwo\ ({\em right}), shown in inverted linear grayscale from -7 to 7 nJy / pixel (-0.003 to 0.003 MJy sr$^{-1}$). Each mosaic consists of 33439 exposures totaling 962.6 hours of observations. Shown in white are the locations of the HUDF/XDF and the two parallel fields.
\label{fig:mosaic}}
\vspace{-2.5cm}
\end{figure*}
}

\def\figcoverage{
\begin{figure*}
\begin{center}
$$\includegraphics[height=9.cm,trim=1.0cm 0 0cm 0]{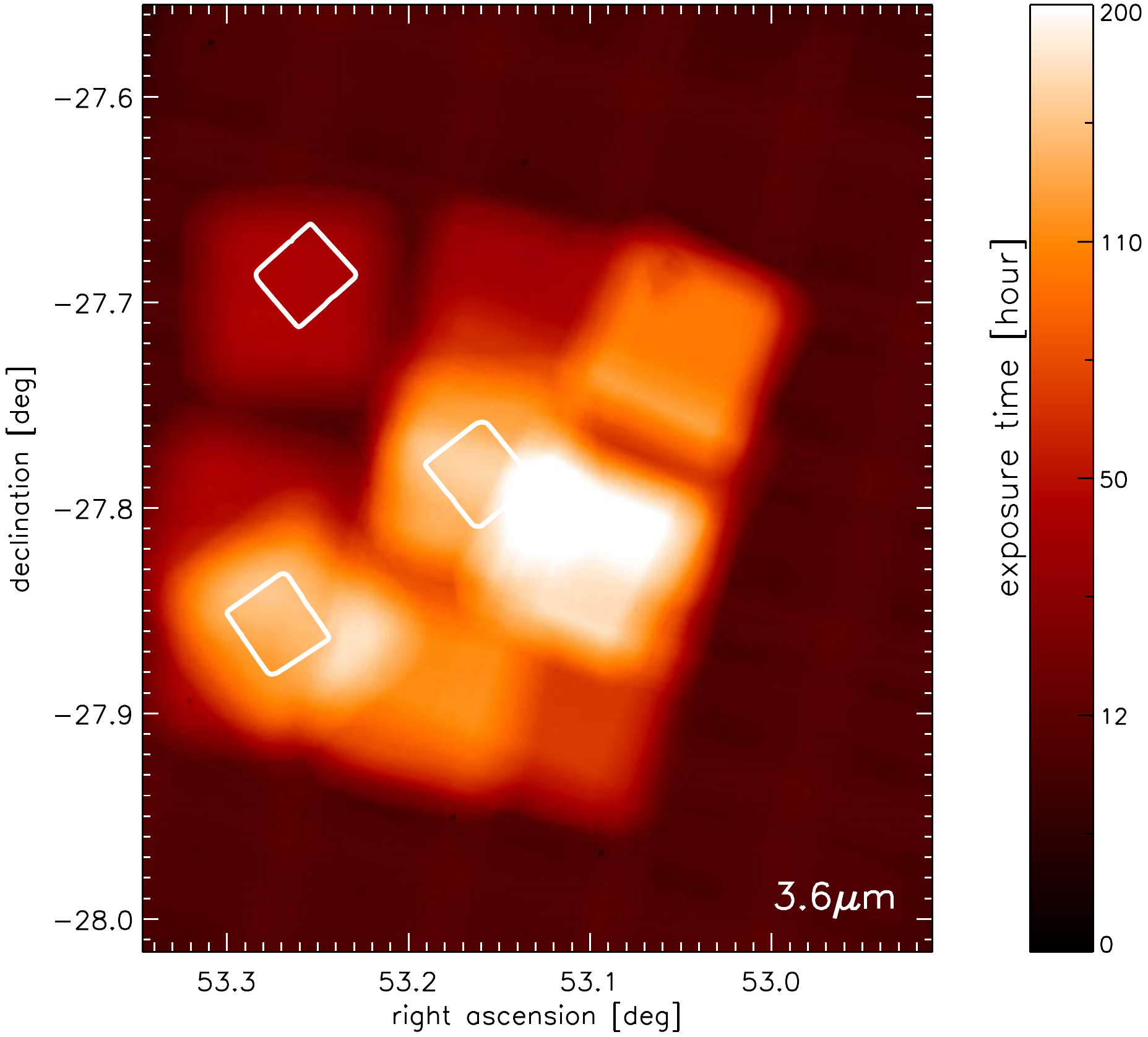}
\includegraphics[height=9.cm,trim=0.0cm 0 0 0]{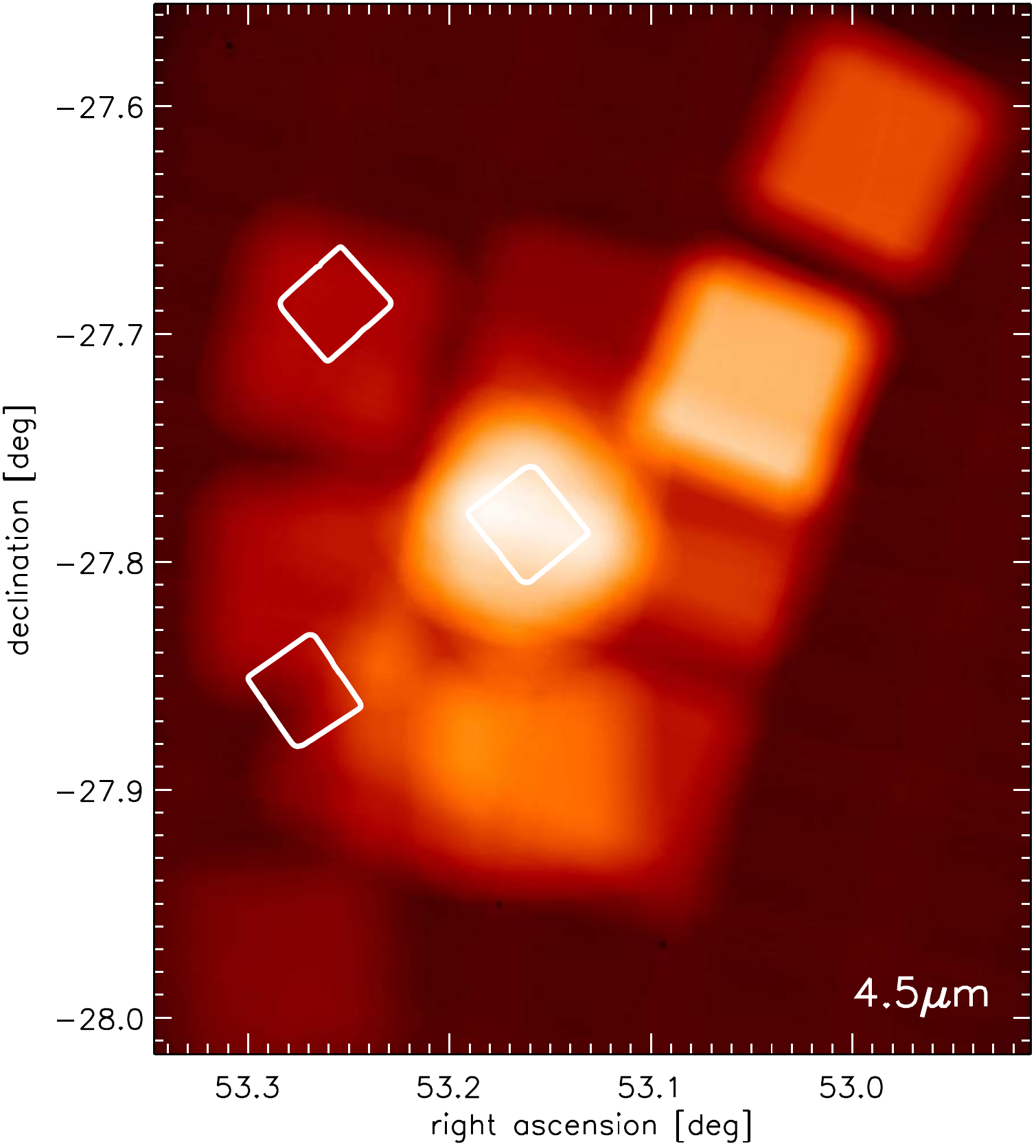}$$
\end{center}
\vspace{-0.2cm}
\caption{
The IRAC coverage maps in GOODS-South and the HUDF fields, shown in heatmap scaling from 0 to 200 hours using a square root stretch. Targeted observations from IUDF and IGOODS and additional fortuitous overlap from many previous IRAC surveys yield total integration time exceeding $>100$ hours over $60$ arcmin$^2$ and $>180-200$ hours over $\sim5-10$ arcmin$^2$.
\label{fig:coverage}}
\end{figure*}
}

\def\figexptime{
\begin{figure}
\begin{center}
\includegraphics[width=8.5cm,trim=0cm 0 0 0]{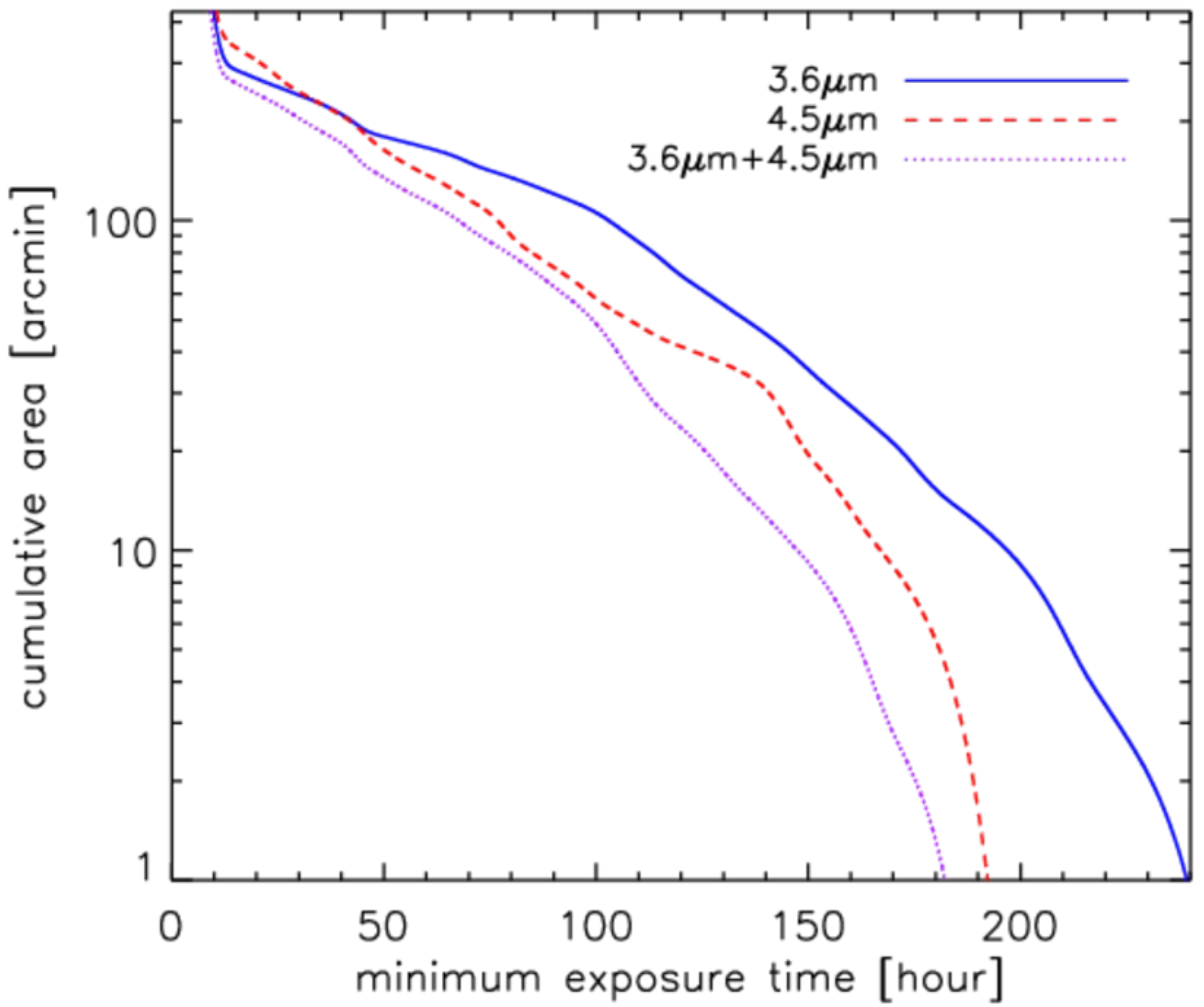}
\end{center}
\vspace{-0.4cm}
\caption{
Area covered versus exposure time for all data over GOODS-South and HUDF fields at \chone\ ({\em blue solid}), \chtwo\ ({\em red dashed}), and joint \chone\ and \chtwo\ ({\em purple dotted}).
The uncoordinated nature of the various programs contributing to the ultradeep mosaics causes the
area covered in both bands to be much smaller than the area covered at $3.6\micron$ or $4.5\micron$. \label{fig:exptime}}
\vspace{0.4cm}
\end{figure}
}

\def\figcol{
\begin{figure*}
\begin{center}
\includegraphics[width=17.cm,trim=0cm 0 0 0]{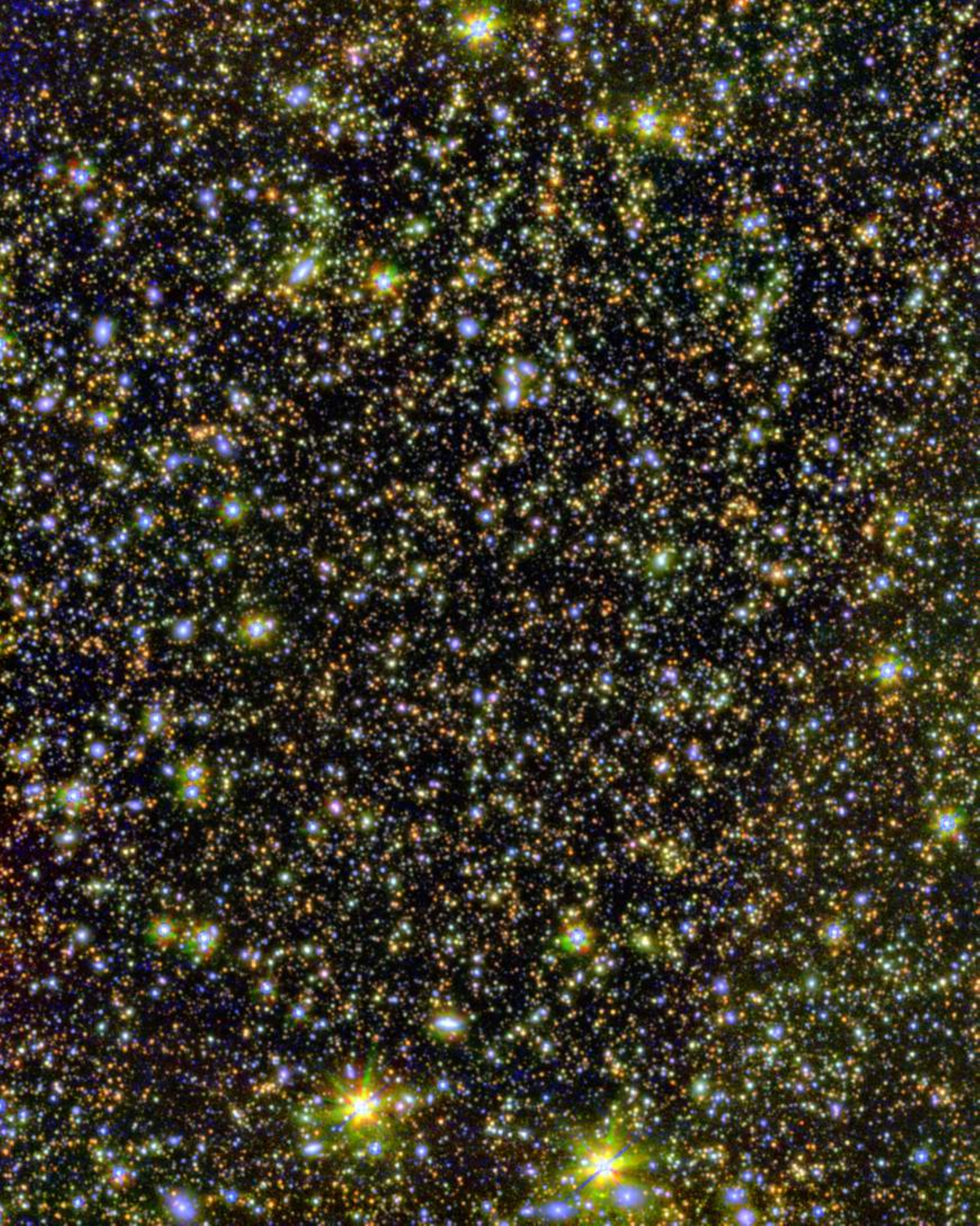}
\end{center}
\vspace{-0.cm}
\caption{
A color composite image of the central deepest region of the GOODS-S field. Deep $K_s-$band data from the TENIS (Hsieh et al. 2012) and HUGS (Fontana et al. 2014) programs are shown as blue, IUDF \chone\ is green, and \chtwo\ is red. The field size is $18\arcmin \times 22\arcmin$ and North up is up.
\label{fig:col}}
\vspace{0.4cm}
\end{figure*}
}

\def\figimprovement{
\begin{figure*}
\leavevmode
\vspace{0.3cm}
\begin{center}
\includegraphics[height=13.5cm,trim=0cm 0 0 0]{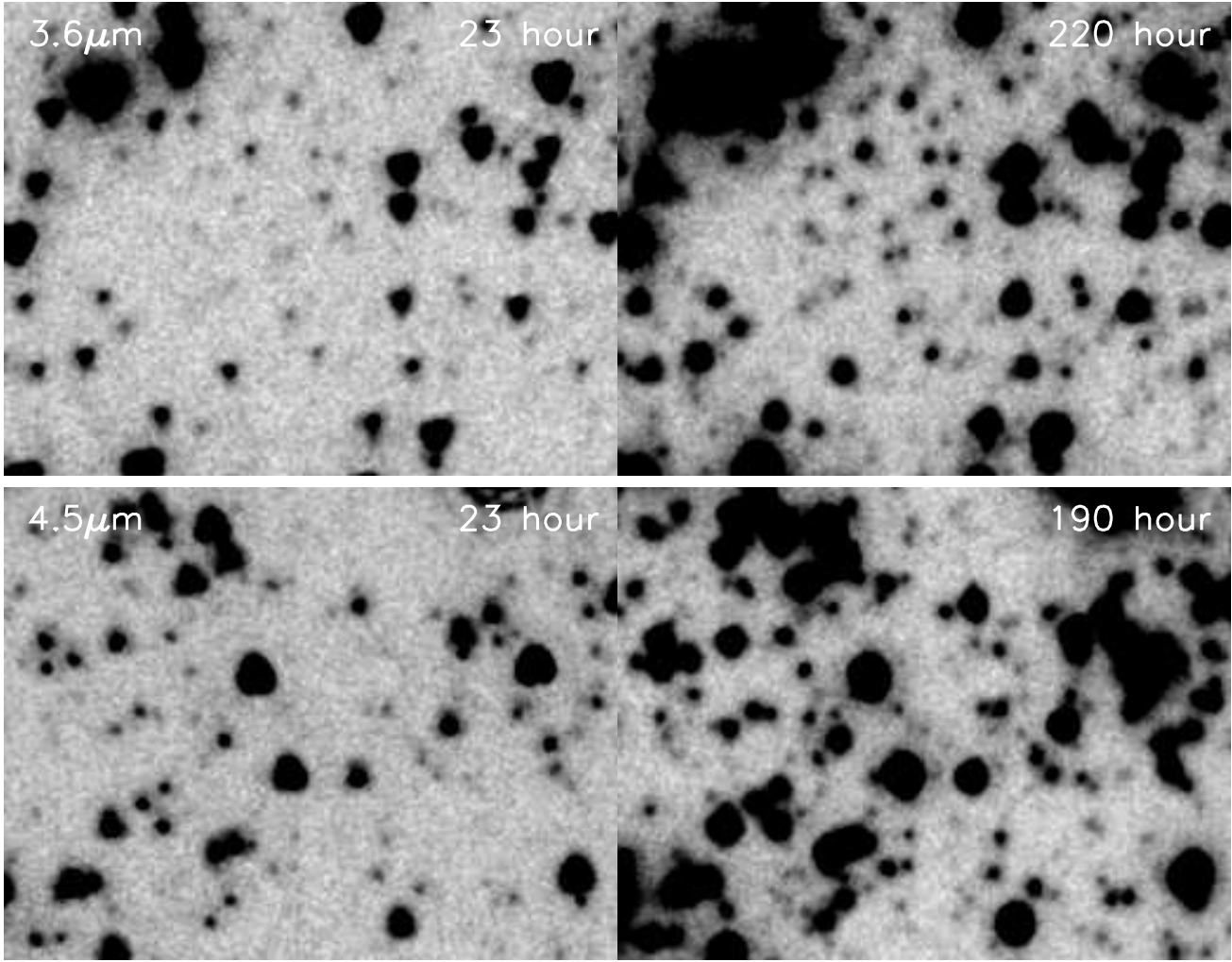}
\end{center}
\vspace{0.0cm}
\caption{
A comparison of Spitzer/IRAC band images of 23 hour  exposure
time (GOODS program single epoch; {\em left}) and the new ultradeep
imaging at $\sim200$ hours of this paper ({\em right}). Different
$1.5\arcmin\times1.0\arcmin$ locations are shown for
\chone\ ({\em top}) and \chtwo\ ({\em bottom}). Image panels are
shown in inverted linear grayscale keeping the background noise at a constant level.
The stretch used is -9 to 9 nJy / pixel (-0.0042 to 0.0042 MJy sr$^{-1}$) at
23 hours and -3 to 3 nJy / pixel (-0.0014 to 0.0014 MJy sr$^{-1}$) at $\sim200$ hours.
A large improvement in signal-to-noise ratio with increased exposure time is
visible and a larger number of faint detected sources.\label{fig:improvement}}
\vspace{0.5cm}
\end{figure*}
}

\def\figdiag{
\begin{figure*}
\leavevmode
\vspace{-0.3cm}
\begin{center}
\includegraphics[width=16.5cm,trim=0cm 3cm 0cm 0cm]{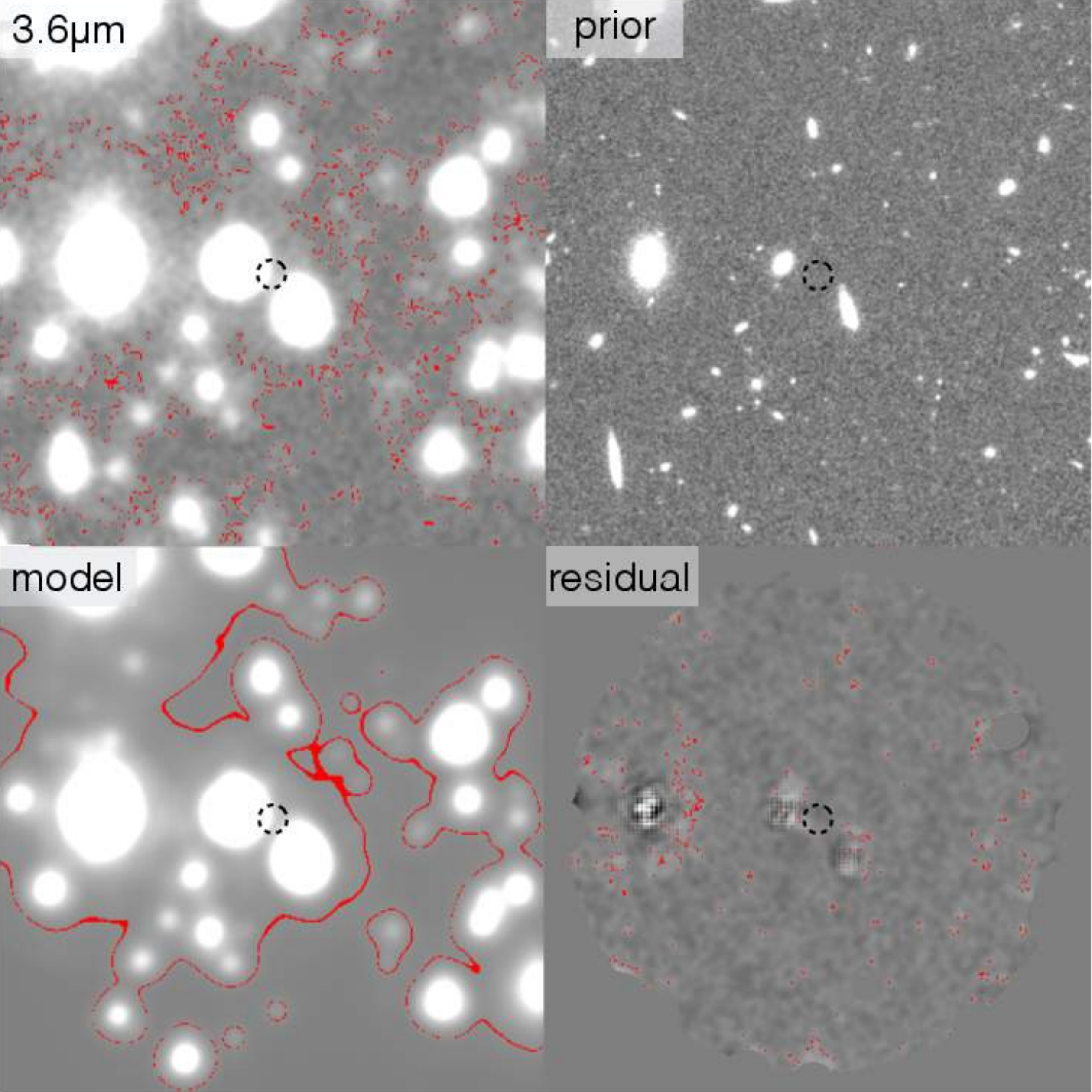}
\end{center}
\vspace{2cm}
\caption{
A demonstration how prior-based IRAC photometry can recover
the full depth of the IRAC data.
({\em top left}) An ultradeep (146 hour exposure time)
$40\arcsec \times 40\arcsec$ section of the IRAC \chone\
mosaic. The red contours shows the $2.5\sigma$ isophote above
the background, indicating that $\sim70\%$ of the background
is contaminated by the PSF wings of sources. The black dashed
aperture shows the location where a flux measurement is desired.
({\em top Right}) Deep HST/WFC3 imaging of the same location on the
sky, which accurately determines the positions and sizes of the sources.
({\em bottom left}) A model is constructed by first convolving
each WFC3 detected source by a kernel to approximate the IRAC PSF,
and then fitting the flux for each individual source
simultaneously. A high quality IRAC PSF model is needed to account for the
PSF wings. ({\em bottom right}) The residual image
shows that the sources are modeled and subtracted very well and that
source confusion is greatly reduced. Small residuals remain
around bright sources due to intrinsic color gradients and small imperfections
in the PSF. The flux measurement in the central aperture
in the residual image is within $1\sigma$ of the background.
\label{fig:diag}}
\vspace{0.3cm}
\end{figure*}
}

\def\fignoise{
\begin{figure*}[t]
\begin{center}
\includegraphics[width=14cm,trim=0cm 0cm 0cm 0cm]{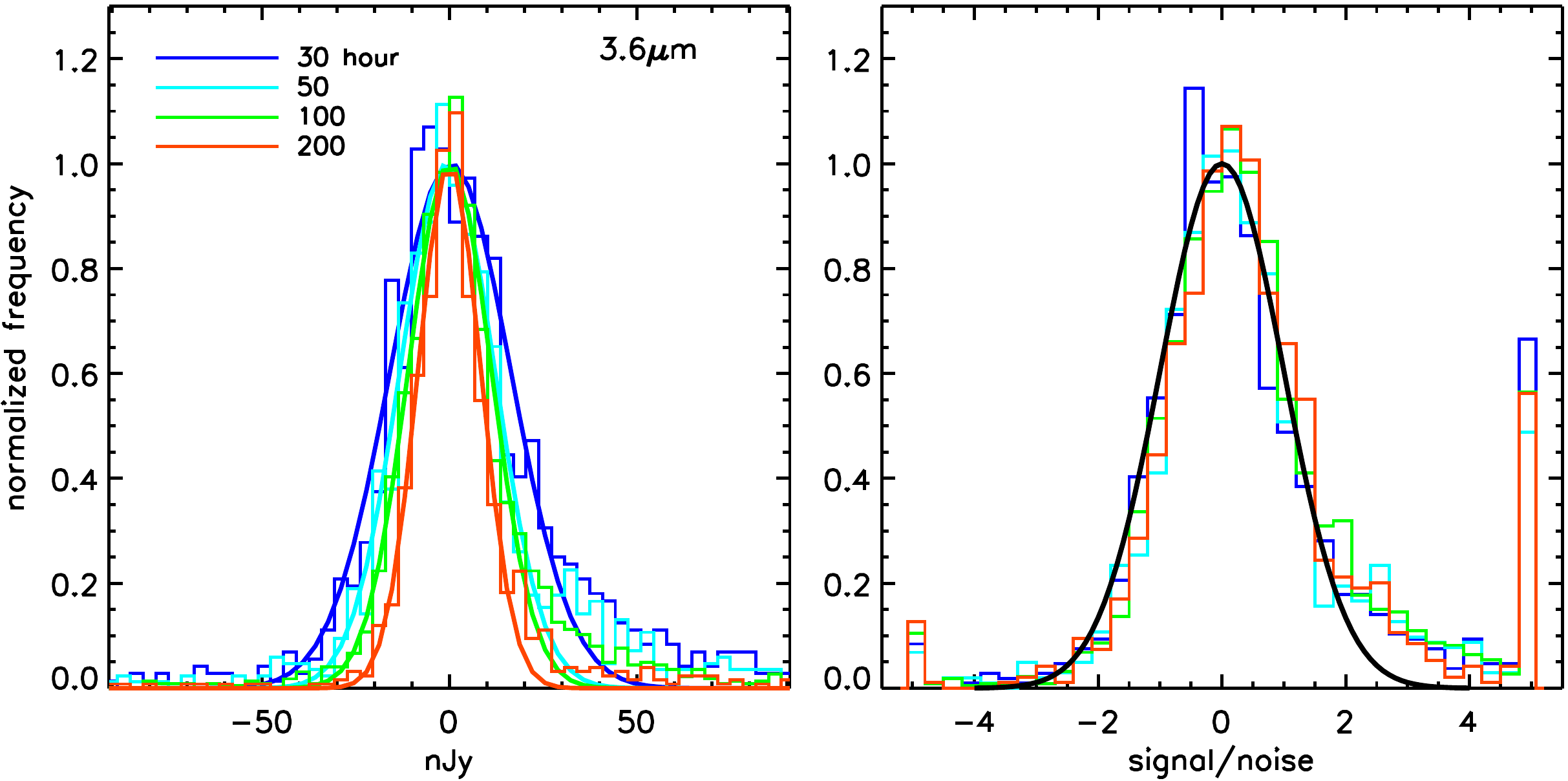}
\includegraphics[width=14cm,trim=0cm 0cm 0cm 0cm]{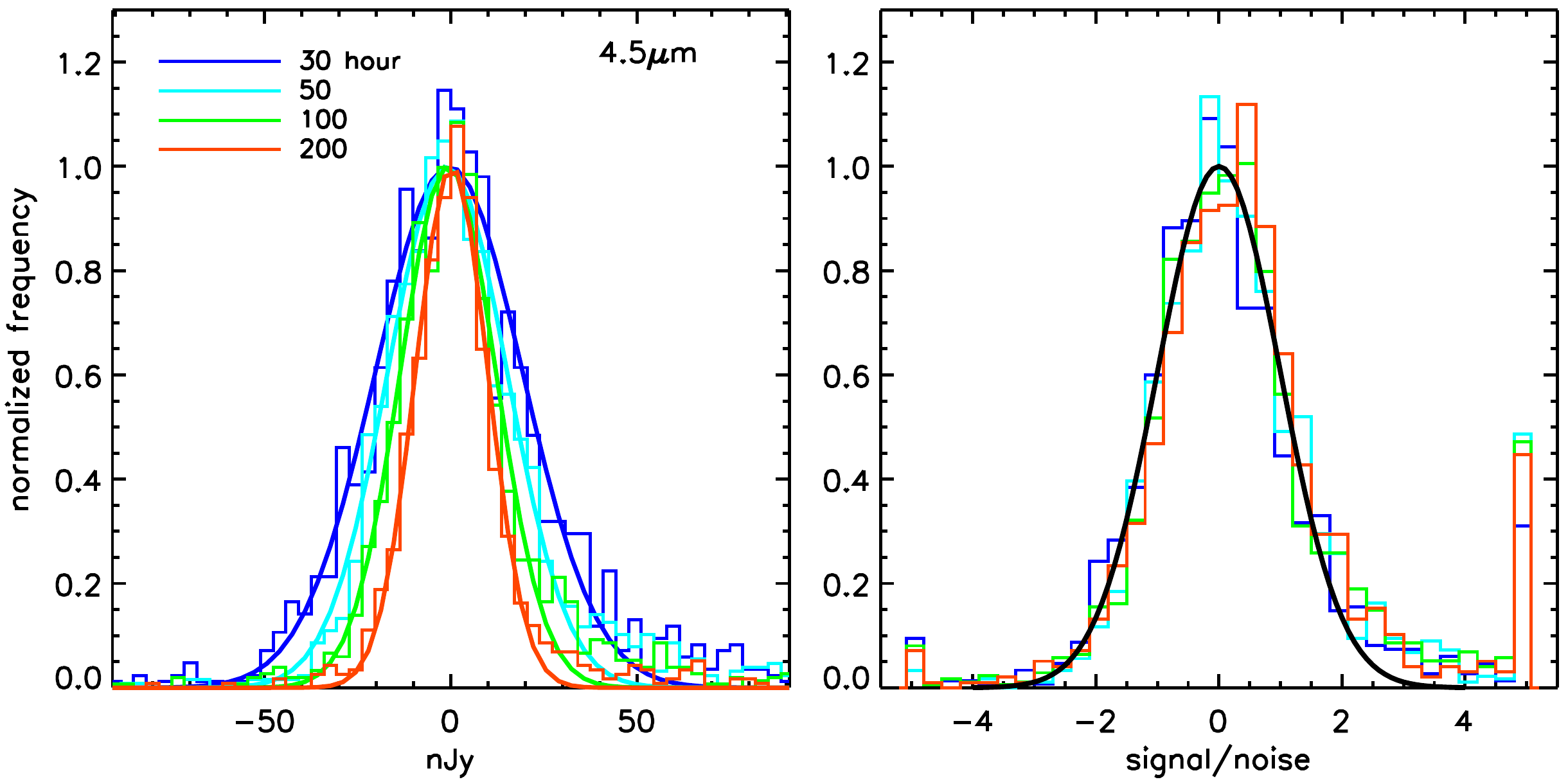}
\end{center}
\vspace{0.cm}
\caption{
({\em left}) Histograms of measured fluxes of artificial sources of zero flux, placed on 15,000 random locations in the full-depth mosaic, and grouped by integration time. The fluxes were measured in circular apertures of $D=2\farcs0$ after modeling and subtracting neighboring sources following the procedure in Fig. \ref{fig:diag}.  The solid lines show
gaussian fits to the histograms. ({\em right}) The histogram of extracted fluxes divided by the local background rms in $1\farcs8\times1\farcs8$ binned pixels. The black curves show
a standard normal \mathN$(0,1)$\ which would be expected in the absence of confusion,
indicating that any residual confusion is not severe for most of the sources, even at 200
hours depth. There is a slight skewness towards positive flux levels, indicated by excess positive residuals for $\sim5\%$ of the sources. About $12\%$ of the fluxes deviate by more than $5\sigma$.
\label{fig:noise}}
\vspace{0.6cm}
\end{figure*}
}

\def\figdepth{
\begin{figure*}
\begin{center}
\includegraphics[height=7.8cm,trim=3.5cm 0 0cm 0]{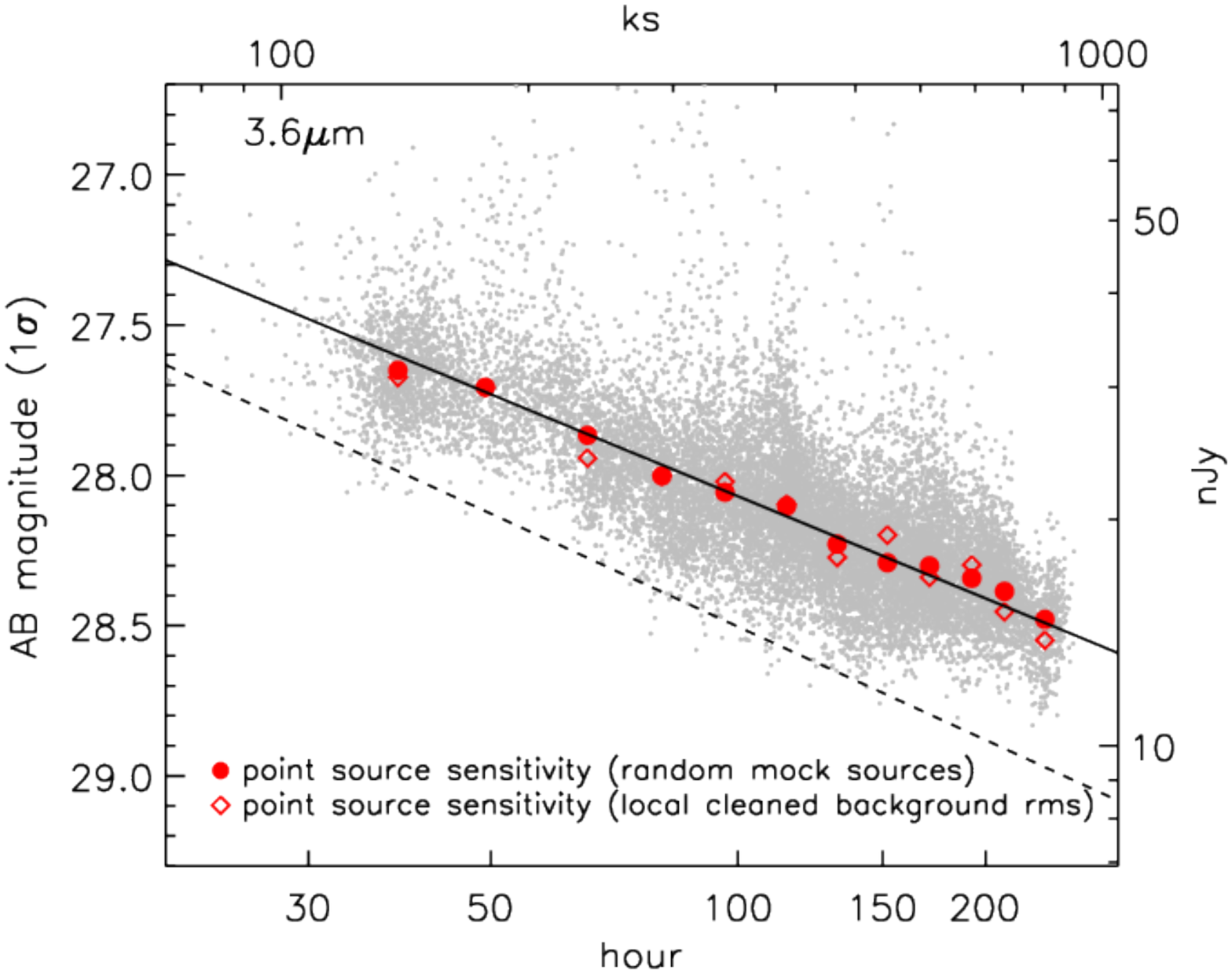}
\includegraphics[height=7.8cm,trim=2.8cm 0cm 0cm 0cm]{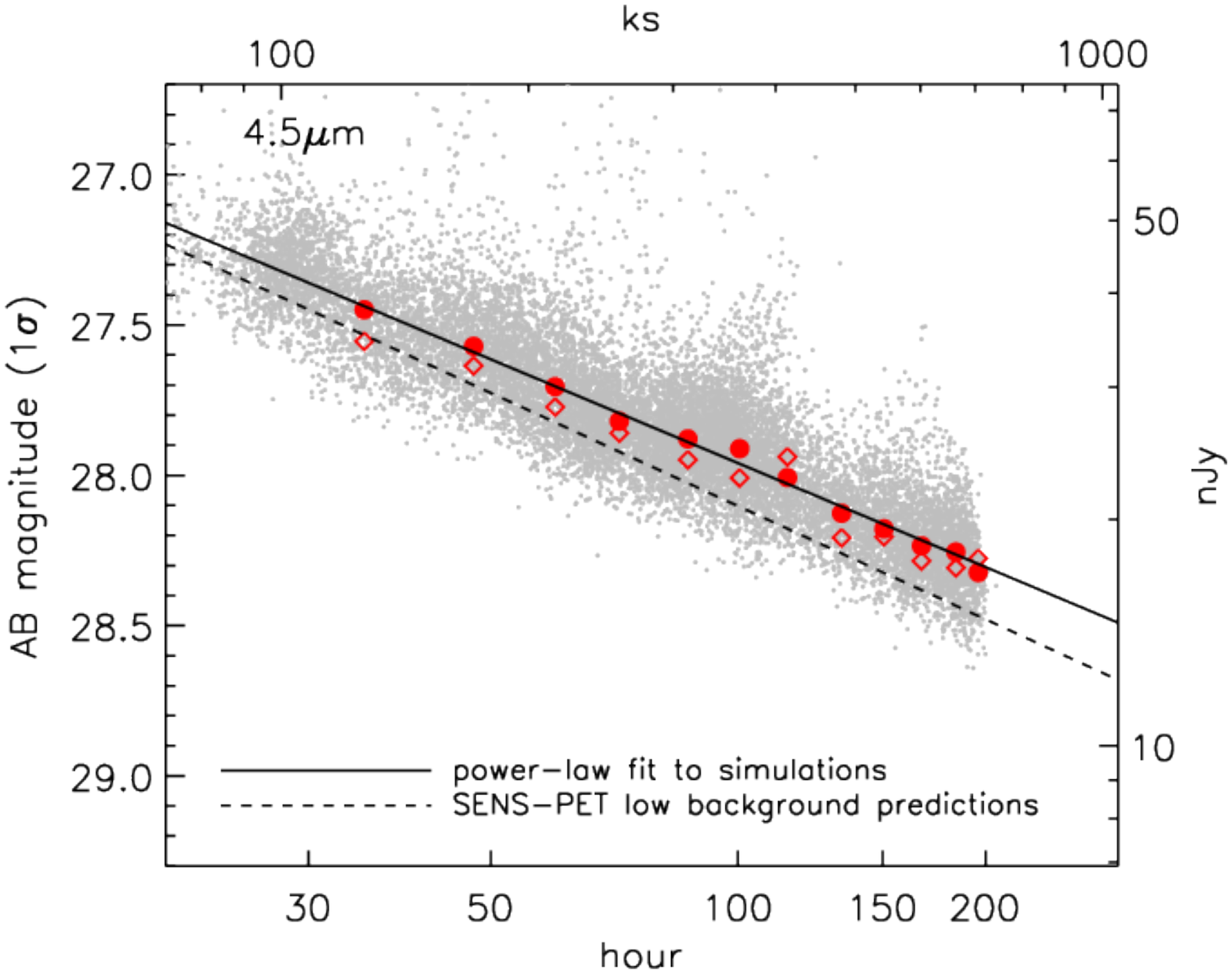}
\end{center}
\vspace{-0.8cm}
\caption{
The relation between median point source sensitivity as a function of integration time based on simulations. Gray points points show the point source fluxes extracted at a large number random locations, after fitting and subtracting neighboring sources using the WFC3 images as a prior.  Red solid points show their
medians in bins of exposure time. Open diamonds show the local background rms away from bright sources. The solid line is a power-law fit to the red solid points, with a best fit slope of $t_{exp}^{-0.45\pm0.01}$ in both IRAC bands. The decrease in noise with exposure time is only slightly slower (at $2\sigma$ significance in each filter) than the $1/\sqrt{t_{exp}}$ expected for Poisson noise, without evidence for a confusion limit or noise floor. The dashed line show predictions from the SENS-PET exposure time calculator.
\label{fig:depth}}
\end{figure*}
}

\def\fighighz{
\begin{figure}
\vspace{0.cm}
\begin{center}
\includegraphics[width=8.cm,trim=0cm 0cm 0cm 0]{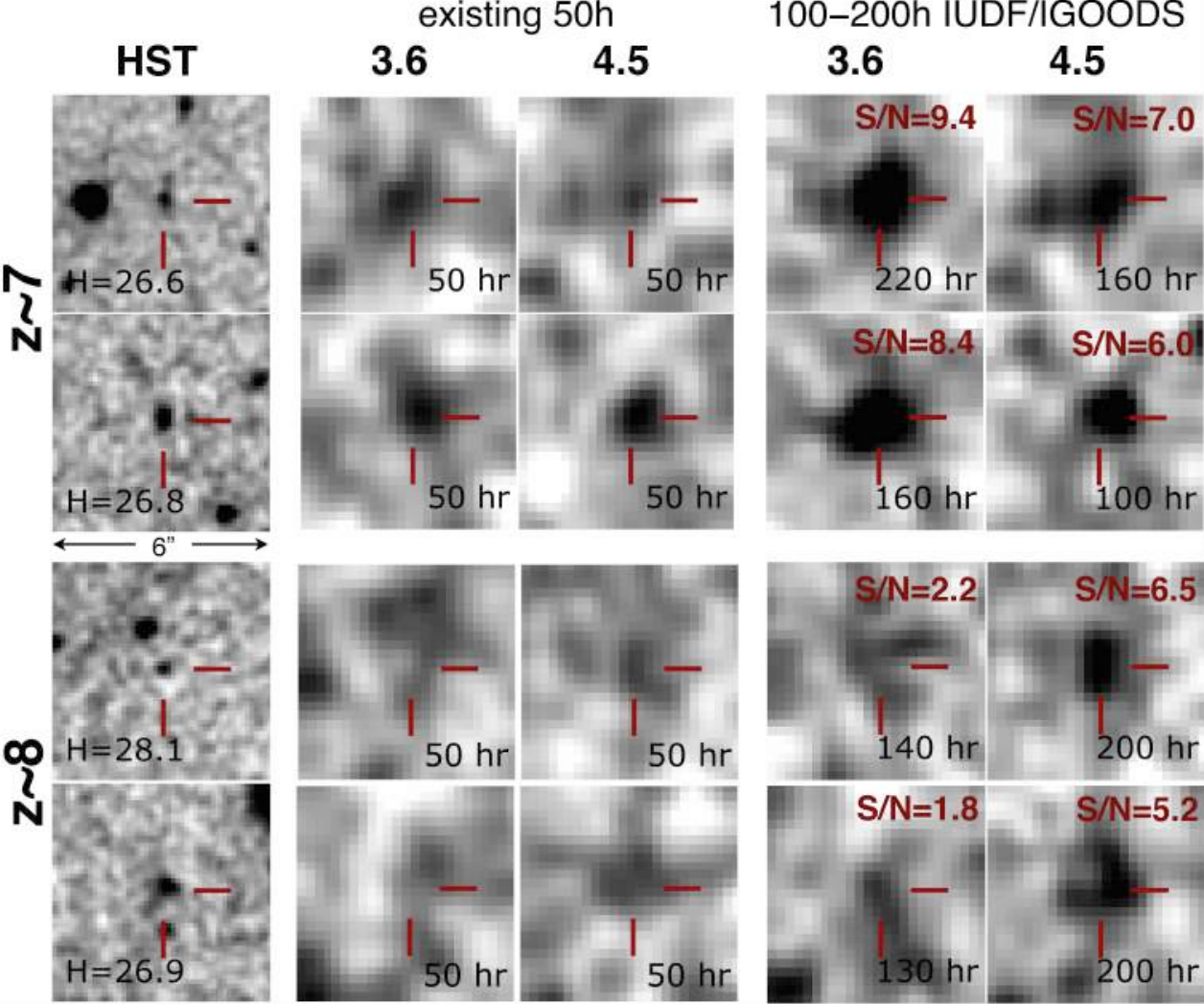}
\end{center}
\vspace{-0.cm}
\caption{Inverted grayscale image stamps of two $z\sim7$ and two $z\sim8$ galaxy candidates in GOODS-South, after modeling and subtracting flux of neighboring sources based on the high-resolution HST image. The panels compare the 50 hour IRAC existing data to the full $100-200$ hour dataset including IUDF + IGOODS (right columns).The stamps are $6\times6\arcsec$. Existing 50 hour data
refer to a combination of GOODS-S (PID 194) and SCANDELS (PID 80217) data. The observed IRAC color changes between $z\sim7$ and $z\sim8$ galaxies (bright at $3.6\micron$ vs bright at $4.5\micron$) as strong [\ion{O}{3}]+$H\beta$ line emission moves from \chone\ to \chtwo e.g., Labb\'e et al. 2013), Sources as faint as ($H_{AB}\sim28$ mag) are detected. \label{fig:highz}}
\vspace{0.5cm}
\end{figure}
}

\def\fighighzpassive{
\begin{figure}
\vspace{0.0cm}
\begin{center}
\includegraphics[width=8.cm,trim=0cm 0.0cm 0.0cm 0.0cm]{z6PassiveExample_small.pdf}
\leavevmode
\vspace{-2cm}
\caption{
Ultradeep IRAC detects objects of unknown origin. These source are undetected in all HST bands in GOODS-South ($<2\sigma$), undetected in the FIR/sub-mm, and are only seen in IRAC at \chtwo\ and \chtwo\ (at $\sim10\sigma$). Their observed SEDs can be fit with quiescent galaxy models at high redshift $z\sim6$ and if this interpretation is correct, they would be the  quenched remnants of starbursts at higher redshifts.
\label{fig:highzpassive}}
\vspace{0.cm}
\end{center}
\end{figure}
}

\title{Ultradeep IRAC imaging over the HUDF and GOODS-South\altaffilmark{1}: \\ survey design and imaging data release}.

\author{I. Labb\'e\altaffilmark{2},
P. A. Oesch\altaffilmark{3,4},
G. D. Illingworth\altaffilmark{5},
P. G. van Dokkum\altaffilmark{4},
R. J. Bouwens\altaffilmark{2},
M. Franx\altaffilmark{2},
C. M. Carollo\altaffilmark{6},
M. Trenti\altaffilmark{7},
B. Holden\altaffilmark{5},
R. Smit\altaffilmark{8},
V. Gonz\'alez\altaffilmark{9},
D. Magee\altaffilmark{5},
M. Stiavelli\altaffilmark{10},
M. Stefanon\altaffilmark{2}
}

\altaffiltext{1}{Based on observations made with the NASA/ESA {\it Hubble Space
Telescope}, obtained from the data archive at the Space Telescope Science
Institute. STScI is operated by the Association of Universities for Research
in Astronomy, Inc. under NASA contract NAS 5-26555. Based on observations made
with the {\it Spitzer Space Telescope}, which is operated by the Jet Propulsion
Laboratory, California Institute of Technology under a contract with NASA.
Support for this work was provided by NASA through an award issued by JPL/Caltech.}
\altaffiltext{2}{Leiden Observatory, Leiden University,P.O. Box 9513, 2300 RA Leiden, The Netherlands}
\altaffiltext{3}{Yale Center for Astronomy and Astrophysics, Physics De- partment, New Haven, CT 06520, USA}
\altaffiltext{4}{Department of Astronomy, Yale University, New Haven, CT 06520}
\altaffiltext{5}{UCO/Lick Observatory, University of California, Santa Cruz, CA 95064}
\altaffiltext{6}{Institute for Astronomy, ETH Zurich, 8092 Zurich, Switzerland}
\altaffiltext{7}{School of Physics, University of Melbourne, Parkville, Victoria, Australia}
\altaffiltext{8}{Centre for Extragalactic Astronomy, Department of Physics, Durham University, South Road, Durham DH1 3LE, UK}
\altaffiltext{9}{University of California, Riverside, 900 University Ave, Riverside, CA 92507, USA}
\altaffiltext{10}{Space Telescope Science Institute, Baltimore, MD 21218, United States}

\begin{abstract}
The IRAC ultradeep field (IUDF) and IRAC Legacy over GOODS (IGOODS) programs are two ultradeep
imaging surveys at \chone\ and \chtwo\ with the {\em Spitzer} Infrared Array Camera (IRAC). The
primary aim is to directly detect the infrared light of reionization epoch galaxies at $z>7$ and to constrain their stellar populations. The observations cover the Hubble Ultra Deep Field (HUDF), including the two HUDF parallel fields, and the CANDELS/GOODS-South, and are combined with archival data from all previous deep programs into one ultradeep dataset. The resulting imaging reaches unprecedented coverage in IRAC \chone\ and \chtwo\ ranging from $>50$ hour over $150$ arcmin$^2$, $>100$ hour over $60$ sq arcmin$^2$, to $\sim200$ hour over $5-10$ arcmin$^2$. This paper presents the survey description, data reduction, and public release of reduced mosaics on the same astrometric system as the CANDELS/GOODS-South WFC3 data. To facilitate prior-based WFC3+IRAC photometry, we introduce a new method to create high signal-to-noise PSFs from the IRAC data and reconstruct the complex spatial variation due to survey geometry. The PSF maps are included in the release, as are registered maps of subsets of the data to enable reliability and variability studies. Simulations show that the noise in the ultradeep IRAC images decreases approximately as the square root of integration time over the range $20-200$ hours, well below the classical confusion limit, reaching $1\sigma$ point source sensitivities as faint as of 15 nJy (28.5 AB) at \chone\ and 18 nJy  (28.3 AB) at \chtwo. The
value of such ultradeep IRAC data is illustrated by direct detections of $z=7-8$ galaxies
as faint as $H_{AB}=28$.
\end{abstract}

\section{Introduction}
Recent years have seen dramatic progress in studies of the early universe, in large part due to sensitive observations with the Wide Field Camera 3 (WFC3) on {\em HST} which detects
the rest-frame UV light of distant galaxies. Studies now routinely identify large numbers of Lyman Break Galaxies (LBGs) in the first billion years of the universe (redshifts $6<z<8$) at the edge of the reionization epoch (e.g., Oesch et al. 2012; McLure et al. 2013; Finkelstein et al. 2012; Grazian et al. 2012, Schmidt et al. 14). Recently, Hubble pushed the frontier even further, finding several galaxies at higher redshifts $z>9$ (around 500 million years after the Big Bang, e.g., Bouwens et al. 2011, Zheng et al. 2012, Ellis et al. 2013, Oesch et al. 2014).

While HST is crucial for selecting the galaxies and determining the redshifts, {\em Spitzer}/IRAC (IRAC; Fazio et al. 2004) excels at detecting the infrared emission of high redshift galaxies. IRAC is currently the only instrument capable of measuring the rest-frame optical light of sources at $4 < z < 10$. The combination of Hubble and Spitzer has proven extremely powerful and provided estimates of the build up of the stellar mass density (e.g., Labb\'e et al. 2010, Gonzalez et al. 2011, Stark et al. 2013, Oesch et al. 2014, \citealt{Duncan14,Grazian14}) and the average specific SFR at $3 < z < 7$ (Gonzalez et al. 2010,2014, Stark et al. 2013, \citealt{Steinhardt14,Salmon14}). Comparing average IRAC colors of redshift $z\sim4-8$ galaxies subsequently showed that star forming galaxies must exhibit very strong nebular emission lines, boosting the Spitzer/IRAC photometry (e.g., \citealt{Sch10}, Labb\'e et al. 2010a,2010b,2013, Shim et al. 2011, Stark et al. 2013, Gonzalez et al. 2014, Smit et al. 2014). This realization has led to the first estimates of nebular emission line equivalent width at $z>4$ and improved estimates of the stellar masses (e.g., Shim et al. 2011, Labb\'e et al. 2013, Stark et al. 2013), which is of vital importance for understanding the mass build up, feedback,  and metal production in the earliest stages of galaxy formation. The current-best example of joint Hubble+Spitzer studies was the robust detection of a small sample of very bright $z\sim10$ candidate galaxies and a first estimate of the galaxy stellar mass density at only 500 Myr after the Big Bang (Oesch et al. 2014). The joint {\it HST}+{\it Spitzer} Frontier Fields campaigns provided other examples of bright, lensed high redshift galaxies (e.g., \citealt{Atek14,Laporte14,Zheng14,Bradac14})

\figlayout

Nevertheless, Spitzer/IRAC observations of earlier programs such as
GOODS (PID 194; PI Dickinson) were only deep enough to individually
detect a small fraction of the $z>6$ sources. For example, Labb\'e et
al. 2010b reported only 2/13 detected at $>5\sigma$ from a sample of
$H_{AB}<27.5$ galaxies at $z\sim7$ over the HUDF. Stacking was necessary to access typical $<L^*$ galaxies (e.g., Labb\'e et al. 2010a) as the \irac\ fluxes of individual sources were too low signal-to-noise (SNR) to be useful. In general, to extract meaningful information from the rest-frame optical SEDs, it is necessary to obtain SNR ratios of $>5$ in each of the $3.6$ and $4.5\mu$m band for typical sources at $z>7$.

\begin{deluxetable*}{lllcrrrr}
\leavevmode
\tablecaption{Summary of IRAC observations}
\tablecolumns{5}
\tablehead{
\colhead{program} & \colhead{PID} & \colhead{PI} & \colhead{max exp.(h)$^a$} & $\#$ pointings & \colhead{total exp.(h)} & \colhead{$\#$ frames} & SSC pipeline version$^d$}
\startdata \\
IUDF      & 70145  & Labb\'e   & 100 & 3 & 215.3   &  8280  & S19.0.0/S18.18.0 \\ \\
IGOODS    &  10076 & Oesch     & 46  & 2 & 65.5    &  2520 &  S19.1.0 \\ \\
GOODS     &  194$^c$   & Dickinson & 46  & 8 & 180.4   &  3356  & S18.25.0 \\ \\
ERS       & 70204 & Fazio      & 75  & 2 & 162.9  &  6264   &  S18.18.0\\ \\
S-CANDELS & 80217  & Fazio     & 25  & 4 & 101.1  & 3888   & S19.0.0/S19.1.0 \\ \\
SEDS      & 60022  & Fazio     & 12  & 20$^b$ & 209.3$^b$ &  8051  & S19.0.0/S18.18.0 \\ \\
UDF2      & 30866$^c$  & Bouwens   & 28.1 & 1 & 28.1 &  1080   & S18.25.0  \\ \\
\tableline \\
total & & & & & 962.6 & 33439  & \\
\add{\tablenotetext{a}{Maximum exposure time per position on the sky per channel.}}
\tablenotetext{b}{Only the central $\sim60$\% of the full SEDS data are used.}
\tablenotetext{c}{Cryogenic mission observations; all other programs are warm mission.}
\tablecomments{Program PID 20708 was omitted because the exposure time is negligible over
the central parts of the GOODS-S region.}
\tablenotetext{d}{The calibration pipelines used were the most recent available from the Spitzer
heritage archive at the time of writing. No significant changes since S18.18.0 have
been reported for 3.6 and 4.5\micron\ observations.}
\label{tab:observations}
\end{deluxetable*}

\begin{deluxetable}{lllcc}
\leavevmode
\tablecaption{Summary of individual AORs}
\tablecolumns{5}
\tablewidth{8cm}
\tablehead{
\colhead{PID} & \colhead{AOR key} & \colhead{MJD$^a$} & \colhead{area$_{50}^b$} & \colhead{$<$exptime$>_{50}^c$}  }
\startdata \\
       70145  &  40849920 & 55487.9259899 &   41.5 &  1.41 \\
       70145  &  40850176 & 55493.6466019 &   36.5 &  1.55 \\
       70145  &  40850432 & 55493.5162173 &   36.1 &  1.57 \\
       70145  &  40850688 & 55611.5557390 &   34.4 &  1.61 \\
       70145  &  40850944 & 55603.7885426 &   36.5 &  1.56 \\
\tablenotetext{a}{Modified Julian Day (JD-2400000.5) in UTC at start of observation}
\tablenotetext{b}{Total area in arcmin$^2$ with $>50\%$ of the maximum exposure time on sky.}
\tablenotetext{c}{Mean exposure time in hour over area$_{50}$.}
\tablecomments{Table 2 is published in its entirety in the electronic edition of ApJS. A portion is shown here for guidance regarding its form and content.}
\label{tab:aor}
\end{deluxetable}

To achieve this we initiated two ultradeep surveys in areas with existing ultradeep ACS+WFC3 data. The first was the cycle 7 IRAC Ultradeep Field (IUDF) program (PI Labb\'e; PID 70145) covering the HUDF/XDF and the two HUDF parallels to $\sim50-100$ hours. The second was the IRAC Legacy over GOODS (IGOODS) program in cycle 10 (PI Oesch; PID 10076), which was aimed at filling out half of the GOODS-South and GOODS-North areas to $\sim200$ hours depth, but which was only $10\%$ completed before being terminated.

This paper described the survey design, data reduction, image quality analysis, and presents the public data release of the IUDF and IGOODS programs, after combining the two ultradeep programs with all archival data over GOODS-South. The paper is structured as follows: \S2 describes the observations, section \S3 summarizes the data reduction and introduces a new technique for creating PSF maps, \S4 describes the resulting ultradeep IRAC mosaics, their properties, and simulations to test prior-based photometry, \S5 discusses the role of IRAC photometry for high redshift galaxies, while a summary is provided in \S6. \ \\

\section{Observations}

The IRAC surveys were all conducted in a single area of the sky, approximately centered on the HUDF in the GOODS-South field around $\alpha = 03:33,\ \delta=-27:48$. This field is
very well suited for IRAC surveys as it has low infrared background and excellent visibility for Spitzer. GOODS-South and the HUDF enjoy the highest quality optical+NIR observations from Hubble (e.g., Giavalisco et al. 2004, Beckwith et al. 2006, Grogin et al. 2011, Koekemoer et al. 2011, Illingworth et al. 2013, Ellis et al. 2013). The high resolution imaging data at shorter wavelengths are necessary for detecting high redshift galaxies and determining their redshift from the location of the redshifted Lyman break. These HUDF data have resulted in some of the largest known samples of high-redshift $z>7$ galaxies. As we shall see, the knowledge of the prior position and size of all sources in the field enables accurate modeling and extraction of the IRAC fluxes.

The GOODS-South field enables the maximum efficiency of any IRAC survey. The existing contiguous WFC3+ACS mosaic over scales of $10-15$ arcmin fills the full IRAC footprint. It also enables parallel \chone\ and \chtwo\ observations, which is relevant as high redshift studies require equally deep observations in both IRAC bands. Finally, very substantial investments in IRAC imaging have already been made in the GOODS fields (amounting to $\gtrsim 500 $ hour per band) so it is
more efficient to continue to build upon previous programs rather than starting from scratch.

Here we combine all programs to create single, contiguous ultra-deep images in the \chone\ and \chtwo\ bands. Below we discuss the individual programs that contributed to the data that were used
to construct the field (dubbed "IRAC Ultra Deep Field", IUDF). \\

\subsection{IRAC Ultradeep Field (IUDF)}
The IUDF cycle 7 program integrated for 210 hours in both IRAC filters, covering the HUDF/XDF
WFC3 field of the HUDF09 survey (PI Illingworth), including its two flanking fields HUDF09-1 and HUDF09-2. These fields are unique due to the concentrated investment of HST time and the large existing samples of $\sim$190 $z>7$ galaxies available immediately for study (Bouwens et al. 2014).

While the HUDF was previously covered with IRAC with 46 hours of cryogenic observations from GOODS (PI Dickinson), the parallel HUDF1 and HUDF2 had received limited and uneven coverage. The IUDF solves this by observing both HUDF parallels to $50-100$ hour at \chone\ and \chtwo, while using
roll angle constraints to obtain deeper imaging on the HUDF/XDF, increasing the exposure time to $100-120$ hour at \chone\ and \chtwo. The HUDF + parallels are the deepest-ever ACS+WFC3+IRAC of any field on the sky. \\

\subsection{IRAC Legacy over GOODS (IGOODS)}

The completion of the IUDF and the success of the first joint ultradeep WFC3+IRAC analyses in the HUDF/XDF (e.g., Oesch et al. 2012,2013, Labb\'e et al. 2013) demonstrated the scientific value of deep IRAC data as well as the feasibility of ultradeep studies. However, much larger samples to even deeper limits are needed for a proper characterization of the $z>7$ universe.

The IGOODS cycle 10 aimed to achieve this by increasing the IRAC depth to a homogenous 200 hours per sky position, while covering much larger areas $\sim 200$ arcmin$^2$ in GOODS-South and GOODS-North. These depths and areas are a sweet spot:  sensitive enough to provide direct detections of sub-L* star forming galaxies at $z\sim8$, while providing enough area for large samples and good statistics ($>200$ galaxies at $z>7$ with $>5\sigma$ IRAC photometry).

Of the approved 800 hours, 200 were earmarked as higher priority to demonstrate the feasibility and usefulness of IRAC data to these limits over the HUDF and GOODS-S. Even though less than 10\%  ($<70$ hour) of the program was executed before the program was terminated due to scheduling conflicts, the program was successful in one aspect. By placing the observations on areas with the deepest overlapping coverage from archival data, it produced the first $>150$ hour deep data in two separate 25 arcmin$^2$ fields in the central part of GOODS-S.\\

\subsection{Archival data}

Apart from the IGOODS and IUDF programs, there exists a wealth of ultradeep IRAC archival data from various programs (most of which are discussed in, e.g., Ashby et al. 2013, Ashby et al. 2015\footnote{We
note that Ashby et al. (2015) present different reductions of very similar observations as described here. We note several key differences: 1) we do not include the shallow and wide field PID 81 and PID 20708 data, but we do include the deep IGOODS PID 10076 observations, 2) reduction and interpolation method are a weighted sum on 0\farcs6 pixel scale in Ashby et al. (2015) versus Drizzling on 0\farcs3 here, and 3) the release in this paper of PSF maps corresponding to the reduced mosaics.}). Table \ref{tab:observations} provides an overview of the programs, the respective PIs, the number of exposures and total integration time. We downloaded all data from the Spitzer Heritage Archive and combined them with our data sets, reducing all in a consistent manner, and coadding them into one ultradeep mosaic. The 7 programs are divided up in 353 Astronomical Observation Requests (AORs), consisting of 33439 exposures, and totaling 3.47 Ms (962.6 hours) in each of the \chone\ and \chtwo\ filters for a total of 1925 hours of IRAC data. At the deepest location the coverage reaches $\sim220$ hours at \chone\ and $\sim190$ hours at \chtwo\ over an area of $\sim5$ arcmin$^2$. \\

\section{Reduction}

The reduction of the IRAC data was carried out starting with
the corrected Basic Calibrated Data (cBCD) generated by the Spitzer
Science Center (SSC) calibration pipeline. A custom
pipeline written by IL was used to post-processes and mosaic the
cBCD frames. The reduction pipeline was also used for reducing
the SIMPLE IRAC Legacy Survey (PI van Dokkum) and described in
detail in Damen et al. 2011.\\

\subsection{IRAC Reduction Process}
The reduction uses a two pass procedure. The first pass comprises
background structure removal, artifact correction, persistence masking,
and a first-pass coaddition. First, a median image is constructed from all
frames in the AOR, to remove background or bias structure and artifacts,
and it is subtracted from each frame\footnote{This procedure works well for
the IUDF, IGOODS, GOODS, and UDF, which take one frame per dither position, but not
for SEDS, S-CANDELS, and ERS, which make use of in-place repeats. This leads to
different bias patterns in the ``first frame'' and the ``repeat frame'' of
each dither. We subtract these by creating two median images, one for all first frames
and one for all repeat frames. }. Then the cBCDs are inspected and
additional artifacts are corrected.  The most important effect is residual
column-pulldown and pull-up. The pull-up/down, caused by bright stars or cosmic
rays at levels $> 10-20$ MJy/Sr in \chone\ and \chtwo, shifts the intensities
of the column above and below in slightly different ways. We correct for it
by subtracting a median above and below the affected pixels after excluding
any sources. Persistence from very bright stars, leaving
positive residuals on subsequent readouts of the array, is masked by rejecting
all highly exposed pixels in the subsequent 4 frames ($\approx400s$).
A constant background pedestal is determined and subtracted from each frame,
 by iteratively masking pixels associated with sources
 and determining the mode of the remaining background pixels.

Finally the post-processed cBCD frames of each AOR are registered and median combined
and a Median Absolute Deviation (MAD) map is calculated (reflecting the uncertainty in
the combined output pixels). The data are very well dithered, hence the images are free
from deviant pixels and can be used to create an object mask.

The second pass comprises cosmic ray rejection, astrometric calibration,
background structure removal, and a final coaddition. First, the first-pass median image is
de-registered and subtracted from each frame. The difference images are divided by
the MAD uncertainty image and used as detection maps for cosmic rays and hot/cold pixels.
Pixels are flagged if they deviate more then 4.5$\sigma_{MAD}$, while pixels adjacent
to outliers are iteratively clipped at a more aggressive $>2.5\sigma_{MAD}$ threshold.
The first-pass image is also used to calibrate the astrometry. The frames in an AORs
are corrected for a simple shift in RA and Dec using sources in common
with the deep WFC3 maps of 3D$-$HST (Skelton et al. 2014). These maps are
convenient as they include the WFC3 observations of the CANDELS/GOODS-South, the WFC3 ERS,
and the HUDF + parallel fields. The rms residuals of individual IRAC
source positions is $0\farcs05-0\farcs07$ rms with systematic
differences on scales of a few arcmin of $\lesssim0\farcs02$. The Skelton et al. (2014)
astrometry was calibrated to the CANDELS/GOODS-South (Koekemoer et al. 2011) mosaics
and to the GEMS (Rix et al. 2004) mosaic for the HUDF parallel fields.

\figpsf
\figpsfmap

A new median background structure map is created from all frames in the AOR,
this time masking objects and outlier pixels. The frames  are then drizzled
(Fruchter \& Hook 2002) per AOR using a pixfrac=0.2 on reference grid defined
by the CANDELS tangent point and a fine $0\farcs3$ pixel scale. A final background
was subtracted by iteratively clipping pixels belonging to objects and subtracting
the mode of the background pixels. Finally, the AORS are weighted
by the exposure time per pixel and combined into the ultradeep mosaic.
The cryogenic observations (GOODS and UDF2) data sets at \chone\ are $\sim$30\%
more sensitive than those of the warm mission hence we increase their contribution
to the final mosaic and exposure time maps by a factor 1.7.
There are no significant differences in sensitivity in \chtwo.
The data release includes the full-depth mosaics in both \chone\ and \chtwo,
as well as mosaics for each AOR in both filters (353 total, on the same grid
and final mosaic position angle). \\

\figmosaic
\figcoverage

\subsection{Point-Spread Function (PSF) Construction}
Accurate point spread functions are needed to facilitate IRAC photometry
using PSF fitting techniques or using the high resolution HST imaging
as a prior. Empirical PSFs created from the reduced mosaics are
preferable, as the observation and reduction processes change the PSFs in subtle ways.
However, extracting clean PSFs to large radii and high dynamic range
is challenging due to crowding of neighboring sources
and the small number of stars usually available in deep blank fields.
To complicate matters, the layout and different rotation angles of the AORs
cause the effective PSF of the combined mosaic to change
rapidly on small spatial scales.

\figcol
\figexptime
\figimprovement

To solve this we generate a spatially varying IRAC PSF. First we take
advantage of the optical stability and the fine sampling to generate
one template ``super PSF'' at \chone\ and \chtwo. Two hundred stars were
identified in deep HST imaging based on their FWHM and magnitude
(e.g., Skelton et al. 2014) and requiring an axis ratio of $b/a>0.85$.
At corresponding locations in each of the 353 AOR mosaics (which are
on the same grid and PA as the full-depth mosaic), image stamps of the
stars were extracted to $R=20\arcsec$ radius. Saturated star images
and those with $SNR<300$ were rejected. The remaining 2050 star images
were then rotated to the native orientation of the IRAC frames to
align the PSF features. Subsequently the images were normalized and
median stacked, sigma clipping outlier pixels due to neighboring
objects. The stacking was iterated three times while growing the
outlier masks by 1 pixel in each iteration. Note that some stars are
imaged in more than 100 distinct AORs. Therefore the distribution of
position angles causes objects close to the stars to fall on different locations on the
IRAC frames. This makes it easier to separate between true PSF
structure and faint signal from neigboring sources, turning the
complex nature of the observations into an asset.

The resulting template PSFs are shown in
Figure \ref{fig:psf} and are of much higher quality and SNR than usual for
deep extragalactic fields. The drizzling on a fine pixel scale of
$0\farcs3$ helps to recover high frequency features of the PSF,
while the large number of high SNR images results in a dynamic range of
$>10,000$.

The second step is to combine the template PSF in such a way that
simulates the combination of the AOR into
the full-depth mosaic. We map the exposure time and rotation
angles of each AOR on a fine grid ($12\arcsec$) covering the output image.
Then we reconstruct the effective full-depth PSF, by rotating\footnote{Rotation and bicubic interpolation of the template PSF introduces very slight smoothing, but is
does not affect the photometry ($<0.2\%$ at small radii).} and weighting the template PSF for each AOR contributing to that grid location.

Figure \ref{fig:psfmap} shows the reconstructed PSFs in
steps of $2.5$ arcmin, illustrating the strong spatial variation. Bootstrap
resampling the star list and repeating the process results in uncertainties
much smaller than the spatial variation in constructed PSF. This indicates
that survey geometry has a much larger impact on the effective IRAC PSF
than the intrinsic variation of the PSF over a single IRAC pointing.
Both the super PSFs and the maps are made available in the data release.

\section{Results}
\subsection{Reduced Image Properties}
The reduced IRAC mosaics are shown in Figure \ref{fig:mosaic} and the
corresponding coverage maps are shown Figure \ref{fig:coverage}. A color
composite using $K_s-$band, \chone\, and \chtwo\ is shown in Figure \ref{fig:col}.
The combined observations of all previous programs results in extremely
deep coverage, due in part to targeted observations over the HUDF/XDF
from the IUDF and IGOODS programs, and in part from fortuitous
overlap from archival data. The uncoordinated nature of the programs
is revealed by the much smaller area covered in both filters
simultaneously: the area is smaller by a factor of $>2$ at $>100$ hr
and factors of $>5$ at $>150$ hours). Simultaneous coverage is
crucial for placing constraints on emission line strengths and
stellar masses at $z>7$ (e.g., Labb\'e et al. 2013). Presently, two small
ultradeep ($180-200$ hr) areas in GOODS-S exist (9 arcmin$^2$ in
\chone\ and \chtwo\ each).

\figdiag

The final mosaics are cosmetically clean and the background
is flat to $5\times10^{-5}$MJy sr$^{-1}$ ($\sim$31 mag/arcsec$^2$ AB)
on scales of $1$ arcmin. The small area that reaches to $180-200$ hours
allows us to evaluate the improvement in background noise relative to the existing deep $25$
hour integrations. As illustrated in Figure \ref{fig:improvement}
the improvement is obvious in both IRAC bands, with large increases in
the number of detected ultrafaint sources and in the SNRs of brighter objects.

The image quality of the full depth mosaics is excellent and constant over the field. The 1-D gaussian full-width at half-maximum (FWHM) over the field
is $1\farcs49\pm0.015$ at \chone\ and $1\farcs48\pm0.025$ at \chtwo. These
values are identical to those of the cryogenic GOODS v0.3 public data release,
and 20\% smaller than those of the SEDS (PID 60022; Ashby et al. 2013) and
SIMPLE mosaics (PID 20708; Damen et al. 2011). The difference with the latter
two programs is due to the native IRAC pixels undersampling the PSF and using
drizzling instead of interpolation when resampling the IRAC frames.

We verify the photometric calibration by comparing the fluxes of bright sources
($<20$ mag AB) in $5\arcsec$ diameter aperture to earlier measurements. The
agreement with the IRAC \chone\ and \chtwo\ imaging of the Spitzer Extended
Deep Survey (SEDS; Ashby et al. 2013) is excellent ($<1\%$ offset).
Comparing to cryogenic GOODS-S imaging (PID 194, PI Dickinson, data release DR3)
reveals that the GOODS fluxes are brighter by 8\% and 2\% in \chone\ and \chtwo\
respectively. This is due to a change in BCD pipeline calibration: the
FLUXCONV values reported in the PID 194 headers (GOODS DR3, v0.30/v0.31,
BCD pipeline S10.5.0) are 7\% and 1\% brighter than the FLUXCONV values in the
most recent calibrations of the same data (BCD pipeline version S18.25.0).
Comparisons to our own reduction of the recalibrated GOODS data shows no offset.

\subsection{Photometry and Confusion}

The total integration times of the mosaics ($50-200$ hours) run well into the classical ``source confusion'' regime for low background extragalactic observations, where crowding by nearby sources affects the reliability of photometry. The classical confusion limit predicted by Franceschini et al. (1991) is $0.6\mu$Jy ($24.5$ AB mag), but in reality confusion is not a hard limit. For example, the classical limit is strictly speaking not relevant when the positions of the sources are known a priori. In GOODS-South and the HUDFs deep ($H_{AB}=27-30$), high-resolution (FWHM$=0\farcs16$) HST/WFC3 imaging is available and the IRAC images are registered to the WFC3 images to very high accuracy ($\lesssim0\farcs02$ systematic). Using the source positions and sizes in the high resolution image, combined with knowledge of the PSFs of WFC3 and IRAC, it is possible extract the source flux by modeling the IRAC surface brightness distribution.
Although surface brightness distribution can vary with wavelength, such procedures already greatly reduce the effect of confusion and open up the possibility of extracting fluxes well beyond the classical limit.

Prior based photometric techniques on blended sources and multi-resolution data sets have been used by many groups in the past with good results (e.g., Fernandez-Soto et al. 1999, Papovich et al. 2001, Shapley et al. 2005, Labb\'e et al. 2005,2006,2010,2013, Grazian et al. 2006, Wuyts et al. 2007, DeSantis et al. 2007, Laidler et al. 2007). As demonstrated in Figure \ref{fig:diag} these techniques can work extremely well.  Note that the photon noise for most sources is negligible compared to the background noise. Therefore, when sources can be modeled and subtracted perfectly, most of the field can be considered empty sky from the perspective of faint source detection.

While good results can already be obtained by simple PSF fitting (i.e., assuming point sources and a negligible size of the high resolution WFC3 PSF), for the best results and smallest residuals near the cores of bright sources, it is necessary to account for both the source size and the detailed shape of the WFC3 and IRAC PSF. This can be done by convolving the isolated high resolution object by a kernel, constructed by deconvolving the low resolution PSF by the high resolution PSF (e.g., Labb\'e et al. 2003, Labb\'e et al. 2005). \\

\fignoise
\figdepth

\subsection{Depth}

The large variation in integration time makes it possible to study the relation between sensitivity and integration time using prior based photometry.  We measure the sensitivity limits of the IRAC images by placing artificial sources of zero flux on 15,0000 random locations in the mosaic and extracting their flux using the WFC3 image as a prior, as previously described and shown in Fig. \ref{fig:diag}. To enable straightforward comparisons with other noise measurements, we do not use the best-fit flux directly but subtract the best-fit model of all neighbors to give a ``cleaned'' image of the source. Then we measure the unweighted flux in $D=2\farcs0$ diameter circular apertures (without further corrections for light outside the aperture).

The histograms of extracted fluxes are shown in Figure \ref{fig:noise}), grouped in bins of integration time. As expected, the scatter histogram becomes progressively narrower with increasing integration time, with no evidence for bias even at the largest integration times.
To compare to the scatter expected from pure background noise, we compute for each fake source the local background RMS in empty regions of the residual image (away from bright sources). We bin by $6\times6$ pixels ($1\farcs8\times1\farcs8$) to approximate the area of a $D=2\farcs0$ aperture. The local empty background RMS is optimistic and only representative of the uncertainty in absence of confusion. As shown in Figure \ref{fig:noise}) ({\em right}) the two estimates agree very well for 90\% of the sources: the histogram of the ratio of aperture flux to local background error resembles a standard normal \mathN$(0,1)$\ distribution. There is a slight skew towards positive flux levels, indicated by excess positive residuals for $\sim5\%$ of the sources in the $2-3\sigma$ range. About $12\%$ of the fluxes deviate by more than $5\sigma$ (10\% high, 2\% low), nearly all due to strong residuals near the centers of very bright IRAC sources. About $3\%$ deviate because of confusion in the high resolution WFC3 prior image.

We further investigate the relationship between contamination fraction and integration time,
defining ``strongly contaminated'' as $>5\sigma$ deviations from the local empty background RMS. Using simple aperture photometry (e.g., SExtractor) on the full-depth mosaics we find high contamination fractions: $\sim80\%$ at \chone\ and $\sim70\%$ in \chtwo. There is only a weak trend of contamination with integration time, likely because most flux comes from moderately bright sources and the PSF surface brightness profile is steep at small radii $R<10\arcsec$ (e.g., Spitzer Observer Manual, SOM, section 6.2.4.1.5). For the cleaned photometry there is no trend with integration time over $20-200$ hour (and a constant $\sim12\%$ contamination). Hence prior based cleaning reduces the contamination fraction for these data sets by a constant factor $6\times$.

Figure \ref{fig:depth} shows the relation between sensitivity and integration time based on the simulated sources. The noise decreases with a power-law slope of $t_{exp}^{-0.45\pm0.01}$ in both IRAC bands. The decrease is only slightly slower (at $2.5\sigma$ significance in each filter) than the $\sqrt{t_{exp}}$ expected for poisson noise. Following the definition of the IRAC integration time calculator (SENS-PET), we convert aperture scatter to point source sensitivity by square root scaling the noise to an equivalent area of $10.5~$arcsec$^2$. This area represents the number of  ``noise pixels'' (see SOM Table 6.1), which would effectively contribute to the uncertainty of linear least-squares fit of a point source. This amounts to optimal weighting by the PSF and improves the SNR by $\sim30\%$ compared to unweighted apertures.\\

The best fit in magnitudes is:
\begin{equation}
 \text{mag}(3.6\mu m,1\sigma,AB) = 25.81 + 1.132 \log_{10}{t_{exp}}
\end{equation}
\begin{equation}
 \text{mag}(4.5\mu m,1\sigma,AB) = 25.66 + 1.141 \log_{10}{t_{exp}}
\end{equation}

or equivalently in flux densities:

\begin{equation}
 \sigma(3.6\mu m,nJy) =   172 * t_{exp}^{-0.453}
\end{equation}
\begin{equation}
 \sigma(4.5\mu m,nJy) =   197 * t_{exp}^{-0.456}
\end{equation}

which gives the median point source sensitivity as function of integration time in hours.
No evidence is found for a confusion limit or noise floor, although the relation is consistently $10-30\%$ less deep than predicted by SENS-PET for low background conditions. A possible explanation for the lower sensitivity is residual confusion by, e.g., sources below our detection limit or a background of faint overlapping PSF wings at larger radii than our PSF model. Note that the true uncertainty for individual sources can be much higher than the median if the source is located close to a bright neighbor. \\

\subsection{Public Data Release}
The data release consists of reduced images of all ultradeep IRAC observations in the GOODS-South. The images are available from the IUDF website\footnote{\url{http://www.strw.leidenuniv.nl/iudf/}} and the Infrared Science Archive\footnote{ \url[http://irsa.ipac.caltech.edu/data/SPITZER/IUDF/]{http://irsa.ipac.caltech.edu/data/SPITZER/IUDF} } (IRSA).\\

The data release contains the following:
\begin{itemize}
\item Science images and exposure time maps in both \chone\ and \chtwo. Our reduction uses the same tangent point as CANDELS on pixel scales of $0\farcs3$, so the IRAC maps can be easily rebinned and registered to HST/WFC3 data.

\item Reduced images of all individual 353 AORs, drizzled onto the same grid, which may be useful to study the reliability or variability of sources.

\item Template PSFs and spatial maps of the weights and position angles of each AOR, allowing the reconstruction of the PSF at arbitrary locations. Example IDL code is provided.
\end{itemize}

The units of the science images are $c $MJy/sr, where constant
c=16.54 represents the change from the native IRAC pixel
scale to 0.3"/pixel due to flux conservation during the reduction process.
Equivalently, flux densities can be obtained by multiplying the image pixel values by
34.994 $\mu$Jy/pixel, corresponding to an image AB zeropoint of 20.04.\\


\section{Examples}

One of the main goals of the IUDF program is to obtain high SNR ($>5\sigma$)
at $3.6$ and $4.5\mu$m for normal $\lesssim L^*$ galaxies in the epoch of reionization.
Comparing the detection rates of $H<27.5$ galaxies at $z>6$ to previous deep IRAC
observations from the GOODS program (PID 194), we find that $\sim46$ hour
GOODS data yields SNR$>5\sigma$ measurements for 25-30\% of the sources,
compared to 75-80\% for $150-200$ hour in the IUDF images.

Here we provide several examples of objects detected in the IUDF images. In Figure \ref{fig:highz}  we show 4 ultrafaint sub-L* galaxies at $z\sim7-8$.  The galaxies are clearly detected at high significance in the new images, compared to the earlier 50 hour deep images. In the deeper images a clear difference in observed IRAC color is seen between the $z\sim7$ and $z\sim8$ galaxies, likely due to  strong [\ion{O}{3}]+$H\beta$ line emission moving from \chone\ to \chtwo\ with increasing redshift. These differences were recently demonstrated in stacked SEDs (e.g., Labb\'e et al. 2013) and in small samples of brighter and lensed galaxies (Smit et al. 2014, Smit et al. 2015), but are now apparent even in individual sub-L* galaxies. This shows the potential of $\sim150-200$ hour data
for placing improved constraints on the emission line strengths of individual galaxies ($H\alpha+$[\ion{N}{2}] at $z=4-5$ and [\ion{O}{3}]+$H\beta$ at $z=7-8$).

\fighighz

Furthermore, ultradeep IRAC data may be the only way to detect potentially important overlooked constituents of the high redshift universe until the arrival of JWST. Massive M$\gtrsim 10^{10}$M$_\odot$ passive galaxies at $z>4$ can be too faint to be detected by Hubble and even actively star forming, dusty galaxies with SFR $50-100$M$_\odot$/yr could have escaped detection by both Hubble and existing FIR/sub-mm surveys at these redshifts. Enigmatic IRAC-selected ``HST-dropouts'' have been identified on the basis of their very red $H-4.5$ colors (e.g., Huang et al. 2011, Caputi et al. 2013). The origin of these objects is unknown as it is difficult to determine their redshifts, but the observed SEDs of some galaxies can be fit with quiescent galaxy models at high redshift $z>4$. If this interpretation is correct, then these objects are the quenched remnants of massive starbursts at earlier times, and they provide compelling targets for early JWST spectroscopic follow up. Such a population likely places powerful constraints on models for star formation quenching, and may inform us indirectly about high mass star formation during the epoch of reionization. \\

\section{Summary}

The IUDF and IGOODS programs are the deepest and most recent probes of the infrared emission at \chone\ and \chtwo\ with Spitzer/IRAC, ideally suited for faint studies of high redshift galaxies. Combining with all ultradeep archival data from all previous programs, and using consistent reduction procedures, we present reduced image mosaics reaching extremely deep coverage of $50-200$ hours and covering all of GOODS-S, the HUDF/XDF, and the two HUDF parallel fields. \\

In summary:
\begin{itemize}
\item We release the full-depth reduced science mosaics at \chone\ and \chtwo\, and the corresponding exposure time maps. The IRAC mosaics are placed on the same astrometric system and reference grid as the CANDELS WFC3 mosaics.\\

\item The combined mosaics are the deepest ever taken at \chone\ and \chtwo\, with the integration times ranging from $>50$ hour over $150$ arcmin$^2$, $>100$ hour over $60$ sq arcmin$^2$, to $\sim180-200$ hour over $5-10$ arcmin$^2$. The image quality is FWHM=$1\farcs49$ in both bands with $<1.5\%$ spatial variation.\\

\item The release also includes the separate reduced mosaics of all individual 353 AORs of the 7 programs involved in this release, registered and drizzled onto the same grid, to study the reliability or variability of sources.\\

\item We present a new procedure to construct IRAC PSF maps from the data, well suited to deep fields with relatively few bright stars and complicated survey geometry with repeat observations onder varying roll angles. The PSF maps are included in the release to facilitate PSF-fitting or joint IRAC+WFC3 photometry.\\

\item Simulations are performed to quantify the confusion due to crowding by neighboring sources.
We demonstrated using the new ultradeep 200 hour data that IRAC observations are not significantly impacted by confusion when using deep high resolution priors from HST/WFC3. In the reduced mosaics $70-80\%$ of the area is originally contaminated by flux of neighboring sources. Using HST-based priors reduces this to a constant $\sim12\%$, with no dependence on exposure time over the range $20-200$ hours. The remaining catastrophic outliers are nearly all very close to the centers of bright IRAC sources and in $3-4\%$ are even confused in the high resolution HST image. In general, prior based photometry works very well, reducing the contamination fraction by $6\times$. \\

\item The simulations further demonstrate that the rms noise in the ultradeep IRAC images decreases nearly as the square root of integration time over the range $20-200$ hours, without any evidence for a hard confusion limit. The maximum  $1\sigma$ point source sensitivities reaches as faint as of 15 nJy (28.5 AB) at \chone\ and 19 nJy  (28.2 AB) at \chtwo. These sensitivities are systematically $10\% - 30\%$ less deep than predicted by the IRAC ETC (SENS-PET), likely due to residual effects of confusion. We provide fitting formulas in \S4.3 to estimate the effective depth as a function of exposure time.\\

\end{itemize}

The value of ultradeep IRAC data is illustrated by direct detections of sub-L* $z>7$ galaxies, where the joint measurement at \chone\ and \chtwo\ places  constraints on the [\ion{O}{3}]+$H\beta$   emission line strengths of individual galaxies to very faint limits $H_{AB}\sim28$. Future observations of larger samples over wider areas will become available as part of Exploration Science program GREATS (GOODS Re−ionization Era wide−Area Treasury from Spitzer, PI Labb\'e), which will map part of GOODS-S and GOODS-N to 200 hours depth. These data offer the prospect of studying the distribution of inferred EWs and comparions to the entire rest-frame SEDs, from HST to ALMA,  will enable studies of the dust attenuation, ionization processes, and star formation histories. The combined HST+Spitzer ultradeep imaging legacy will be useful for planning efficient imaging and spectroscopic follow-up surveys with JWST and provide interesting targets for the first cycles of JWST NIRSPEC observations. Spitzer's heritage will extend well into the JWST era.

\acknowledgements

We thank the anonymous referee for careful comments which improved the clarity of
the paper. We thank Justin Howell for IPAC assistance with the data release.
Support for this work was provided ERC grant HIGHZ no. 227749 and NL-NWO Spinoza.
We acknowledge support from NASA grants NAG5-7697 and HST-GO-11563.
This research has made use of the NASA/ IPAC Infrared
Science Archive, which is operated by the Jet Propulsion Laboratory,
California Institute of Technology, under contract with the National
Aeronautics and Space Administration.

\end{document}